\documentclass[10pt,conference]{IEEEtran}
\IEEEoverridecommandlockouts
\usepackage{cite}
\usepackage{hyperref}
\usepackage{amsmath,amssymb,amsfonts}
\usepackage{graphicx,caption}
\usepackage{cleveref}
\usepackage{booktabs}
\usepackage{tabularx}
\usepackage{textcomp}
\usepackage{xcolor}
\usepackage{xspace}
\usepackage[most]{tcolorbox}
\usepackage{booktabs}
\usepackage{listings}
\usepackage{tablefootnote}
\usepackage[nofillcomment,vlined,ruled,linesnumbered]{algorithm2e}
\usepackage{upquote}
\usepackage{amsmath}
\usepackage{subcaption}
\usepackage[T1]{fontenc}
\usepackage[scaled=0.85]{inconsolata} % use a better font for listings
\PassOptionsToPackage{svgnames}{xcolor}

\definecolor{codegreen}{rgb}{0,0.6,0}
\definecolor{white}{rgb}{1,1,1}
\definecolor{codegray}{rgb}{0.5,0.5,0.5}
\definecolor{backcolour}{rgb}{0.95,0.95,0.92}
\definecolor{codepurple}{rgb}{0.5,0.0,0.5}

\newcommand{\name}{{\textsc{Libro}}\xspace}
\def\BibTeX{{\rm B\kern-.05em{\sc i\kern-.025em b}\kern-.08em
    T\kern-.1667em\lower.7ex\hbox{E}\kern-.125emX}}

\newcommand{\code}[1]{\lstinline[basicstyle=\normalsize\ttfamily]+#1+}
\newcommand{\codealgo}[1]{{\lstinline[basicstyle=\footnotesize\ttfamily]+#1+}}
\begin{document}

\lstset{
    % basicstyle=\ttfamily\footnotesize,
    % breakatwhitespace=false,         
    breaklines=true,
    xleftmargin=1em,                  
    showtabs=false,                  
    tabsize=2,
    numberstyle=\tiny\color{codegray},
    backgroundcolor=\color{white},   
    commentstyle=\color{codegreen},
    keywordstyle=\color{blue},
    stringstyle=\color{codepurple},
    breaklines=true,
    columns=flexible,
    numbers=left,
}

\title{Large Language Models are Few-shot Testers:\\
Exploring LLM-based General Bug Reproduction
}

\author{\IEEEauthorblockN{Sungmin Kang\IEEEauthorrefmark{1}}
\IEEEauthorblockA{\textit{School of Computing} \\
\textit{KAIST}\\
Daejeon, Republic of Korea \\
sungmin.kang@kaist.ac.kr}
\and
\IEEEauthorblockN{Juyeon Yoon\IEEEauthorrefmark{1}}
\IEEEauthorblockA{\textit{School of Computing} \\
\textit{KAIST}\\
Daejeon, Republic of Korea \\
juyeon.yoon@kaist.ac.kr}
\and
\IEEEauthorblockN{Shin Yoo}
\IEEEauthorblockA{\textit{School of Computing} \\
\textit{KAIST}\\
Daejeon, Republic of Korea \\
shin.yoo@kaist.ac.kr}
\thanks{$^{*}$: these authors contributed equally.}
}

\maketitle

\begin{abstract}
Many automated test generation techniques have been developed to aid developers 
with writing tests. To facilitate full automation, most 
existing techniques aim to either increase coverage, or generate exploratory inputs. 
However, existing test generation techniques
largely fall short of achieving more semantic objectives, such as generating 
tests to reproduce a given bug report.
Reproducing bugs is nonetheless important, as our 
empirical study shows that the number of tests added in open source repositories due to issues 
was about 28\% of the corresponding project test suite size.
Meanwhile, due to the difficulties of 
transforming the expected program semantics in bug reports 
into test oracles, existing failure reproduction techniques tend to deal exclusively with 
program crashes, a small subset of all bug reports. 
To automate test generation from general bug reports, we propose \name, 
a framework that uses Large Language 
Models (LLMs), which have been shown to be capable of performing 
code-related tasks.
Since LLMs themselves cannot execute the
target buggy code, we focus on post-processing steps that help us discern when
LLMs are effective, and rank the produced tests according to their 
validity. Our evaluation of \name shows that, on the 
widely studied Defects4J benchmark, \name can generate failure reproducing test 
cases for 33\% of all studied cases (251 out of 750), while suggesting a bug reproducing
test in first place for 149 bugs. To mitigate data 
contamination (i.e., the possibility of the LLM simply remembering the test code 
either partially or in whole), we also evaluate \name against 31 bug reports 
submitted after the collection of the LLM training data terminated: \name 
produces bug reproducing tests for 32\% of the studied bug 
reports.
Overall, our results show \name has the potential to 
significantly enhance developer efficiency by automatically generating tests 
from bug reports.
\end{abstract}


\section{Introduction}
\label{sec:intro}

Software testing is the practice of confirming that software meets specification
criteria by executing tests on the software under test (SUT). Due to the 
importance and safety-critical nature of many software projects, software 
testing is one of the most important practices in the software development 
process. Despite this, it is widely acknowledged that software testing is 
tedious due to the significant human effort 
required~\cite{Haas2021ManualTesting}.
To fill this gap, automated test generation techniques have been studied for
almost half a century~\cite{Miller:1976ht}, resulting in a number of 
tools~\cite{Pacheco:2007oq, Fraser:2013vn} that use implicit oracles 
(regressions or crash detection) to guide the automated process. They are 
useful when new features are being added, as they can generate novel tests with 
high coverage for a focal class.

However, not all tests are added immediately along with their focal class. 
In fact, we find that a significant number of tests originate from \emph{bug 
reports}, i.e., are created in order to prevent future regressions for the bug 
reported. This suggests that \emph{the generation of bug reproducing tests from 
bug reports} is an under-appreciated yet impactful way of automatically writing 
tests for developers. Our claim is based on the analysis of a sample of 300 
open source projects using JUnit: the number of tests added as a result of bug 
reports was on median 28\% of the size of the overall test suite. Thus, the bug 
report-to-test problem is regularly dealt with by developers, and a problem in 
which an automated technique could provide significant help. Previous work in bug 
reproduction mostly deals with crashes~\cite{Soltani2018aa, Nayrolles2015Jcharming};
as many bug reports deal with semantic issues, their scope is limited in practice.

The general report-to-test problem is of significant importance to 
the software engineering community, as solving this problem would allow developers 
use a greater number of automated debugging techniques, 
equipped with test cases that reproduce the reported bug. Koyuncu et 
al.~\cite{Koyuncu2019ifixr} note that in the widely used Defects4J~\cite{
Just:2014aa} bug benchmark, bug-revealing tests \textit{did not exist} prior to 
the bug report being filed in 96\% of the cases. Consequently, it may be 
difficult to utilize the state-of-the-art automated debugging techniques, which 
are often evaluated on Defects4J, when a bug is first reported because they 
rely on bug reproducing tests~\cite{Xiong2017ACS, Li2019DeepFL}. Conversely, 
alongside a technique that automatically generates bug-revealing tests, a wide 
range of automated debugging techniques would become usable.

As an initial attempt to solve this problem, we propose prompting Large 
Language Models (LLMs) to generate tests. Our use of LLMs is based on 
their impressive performance on a wide range of natural language processing 
tasks~\cite{brown2020language} and programming tasks~\cite{Chen2022iw}. 
In this work, we explore whether their 
capabilities can be extended to generating test cases from bug reports. More 
importantly, we argue that the performance of LLMs when applied to this problem 
has to be studied along with the issue of \emph{when} we can rely on the tests 
that LLMs produce. Such questions are crucial for actual developer use: 
Sarkar et al.~\cite{Sarkar2022PwAI} provide relevant examples, showing that 
developers struggle to understand when LLMs will do their bidding when used for 
code generation. To fill this gap of knowledge, we propose \name (LLM Induced 
Bug ReprOduction), a framework that prompts the OpenAI LLM, Codex~\cite{Chen2021ec}, to generate 
tests, processes the results, and suggests solutions only when we can be 
reasonably confident that bug reproduction has succeeded.

We perform extensive empirical experiments on both the Defects4J benchmark and 
a new report-test dataset that we have constructed, aiming to identify the 
features that can indicate successful bug reproduction by \name. We find that, 
for the Defects4J benchmark, \name can generate at least one 
bug reproducing test for 251 bugs, or 33.5\% of all studied bugs from their bug 
reports. \name also successfully deduced which of its bug reproducing attempts
were successful with 71.4\% accuracy, and produced an actual bug reproducing test
as its first suggestion for 149 bugs.
For further validation, we evaluate \name on a recent bug report 
dataset that we built, finding that we could reproduce 32.2\% of bugs in this 
distinct dataset as well, and verifying that our test suggesting heuristics
work in this different dataset as well.

In summary, our contributions are as follows:
\begin{itemize}
    \item We perform an analysis of open source repositories to verify the importance of generating bug reproducing test cases from bug reports;
    \item We propose a framework to harness an LLM to reproduce bugs, and suggest generated tests to the
    developer only when the results are reliable;
    \item We perform extensive empirical analyses on two datasets, suggesting that the patterns we
    find, and thus the performance of \name, are robust.
\end{itemize}

The remainder of the paper is organized as follows. We motivate our research in \Cref{sec:motivation}. Based on this, we describe our approach
in \Cref{sec:approach}. Evaluation settings and research questions are in \Cref{sec:evaluation}
and \Cref{sec:rqs}, respectively. Results are presented in \Cref{sec:results}, while threats to validity are discussed in \Cref{sec:threats}. \Cref{sec:relwork} gives an overview of the
relevant literature, and \Cref{sec:conclusion} concludes.

\section{Motivation}
\label{sec:motivation}

As described in the previous section, the importance of the report-to-test 
problem rests on two observations. The first is that bug-revealing tests are
rarely available when a bug report is filed, unlike what automated debugging 
techniques often assume~\cite{Xiong2017ACS, Li2019DeepFL}. Koyuncu et al.~\cite{
Koyuncu2019ifixr} report that Spectrum-Based Fault Localization (SBFL) 
techniques cannot locate the bug at the time of being reported in 95\% of the 
cases they analyzed, and thus propose a completely static automated debugging 
technique. However, as Le et al.~\cite{Le:2015aa} demonstrate, using dynamic 
information often leads to more precise localization results. As such, a 
report-to-test technique could \emph{enhance the practicality and/or performance of a large portion of the automated debugging literature}.

The other observation is that the report-to-test problem is a perhaps 
underappreciated yet nonetheless important and recurring part of testing.
Existing surveys of developers reveal that developers consider generating tests from bug reports to be a way to improve automated testing. Daka and 
Fraser~\cite{Data2014UTSurvey} survey 225 software developers and point out ways
in which automated test generation could help developers, three of which (deciding what to test, 
realism, deciding what to check) can be resolved using bug reports, as bug reproduction 
is a relatively well-defined activity. 
Kochhar et al.~\cite{
Kochhar2019Practitioner} explicitly ask hundreds of developers on whether they 
agree to the statement ``during maintenance, when a bug is fixed, it is good to 
add a test case that covers it'', and find a strong average agreement of 4.4 on 
a Likert scale of 5.

To further verify that developers regularly deal with the report-to-test 
problem, we analyze the number of test additions that can be attributed to 
a bug report, by mining hundreds of open-source Java repositories. We start 
with the \texttt{Java-med} dataset from Alon et al.~\cite{Alon2019ty},
which consists of 1000 top-starred Java projects from GitHub. From the list of 
commits in each repository, we check (i) whether the commit adds a test, and 
(ii) whether the commit is linked to an issue. To determine whether a commit 
adds a test, we check that its diff adds the \texttt{@Test} decorator along 
with a test body. In addition, we link a commit to a bug report (or an 
\emph{issue} in GitHub) if (i) the commit message mentions 
"(fixes/resolves/closes) \#NUM", or (ii) the commit message mentions a pull 
request, which in turn mentions an issue. We compare the number of tests added 
by such report-related commits to the size of the current (August 2022) test 
suite to estimate the prevalence of such tests. As different repositories
have different issue-handling practices, we filter out repositories that have
no issue-related commits that add tests, as this indicates a different
bug handling practice (e.g. \texttt{google/guava}). Accordingly, we analyze
300 repositories, as shown in \Cref{tab:repo_categories}.

\begin{table}[ht]
    \centering
    \caption{Analyzed repository characteristics\label{tab:repo_categories}}
    \scalebox{0.9}{
    \begin{tabular}{lc}
    \toprule
    Repository Characteristic & \# Repositories\\\midrule
    Could be cloned & 970 \\
    Had a JUnit test (\texttt{@Test} is found in repository) & 550 \\
    Had issue-referencing commit that added test & 300 \\
    \bottomrule
    \end{tabular}
    }
\end{table}

We find that the median ratio of tests added by issue-referencing commits in those 300 repositories, relative to the current test suite size, is 
28.4\%, suggesting that a significant number of tests are added due to bug 
reports. We note that this does not mean 28.4\% of tests in a test suite 
originate from bug reports, as we do not track what happens to tests after they are added.
Nonetheless, it indicates the report-to-test activity plays a significant role 
in the evolution of test suites.
Based on this result, we conclude that the report-to-test generation problem is 
regularly handled by open source developers. By extension, an automated 
report-to-test technique that suggests and/or automatically commits
confirmed tests would help developers in their natural workflow.

Despite the importance of the problem, its general form remains a difficult problem to solve. 
Existing work attempts to solve special cases of the problem by 
focusing on different aspects: some classify the sentences of a 
report into categories like observed or expected behavior~\cite{Song2020Bee}, 
while others only reproduce crashes (crash reproduction)~\cite{Soltani2020aa, 
Nayrolles2015Jcharming}. We observe that solving this problem requires good 
understanding of both natural and programming language, not to mention 
capabilities to perform deduction. For example, the bug report in 
\Cref{tab:ex-report-math-63} does not explicitly specify any code, but a fluent 
user in English and Java would be capable of deducing that when both arguments
are NaN, the `equals' methods in `MathUtils' should return \code{false}. 

One promising solution is to harness the capabilities of pre-trained Large 
Language Models (LLMs). LLMs are generally Transformer-based neural 
networks~\cite{Sarkar2022PwAI} trained with the language modeling 
objective, i.e. predicting the next token based on preceding context. One of 
their main novelties is that they can perform tasks without training: simply by 
`asking' the LLM to perform a task via a textual prompt, the LLM is often 
capable of actually performing the task~\cite{brown2020language}.
Thus, one point of curiosity is how many bugs LLMs can reproduce given a report.
On the other hand, of practical importance is to be able to know \emph{when} we 
should believe and use the LLM results, as noted in the introduction. To this 
end, we focus on finding heuristics indicative of high precision, and minimize 
the hassle that a developer would have to deal with when using \name.% it comes to automatically generated solutions.

\section{Approach}
\label{sec:approach}

\begin{figure}[h!]
\centering
\includegraphics[width=1.0\linewidth]{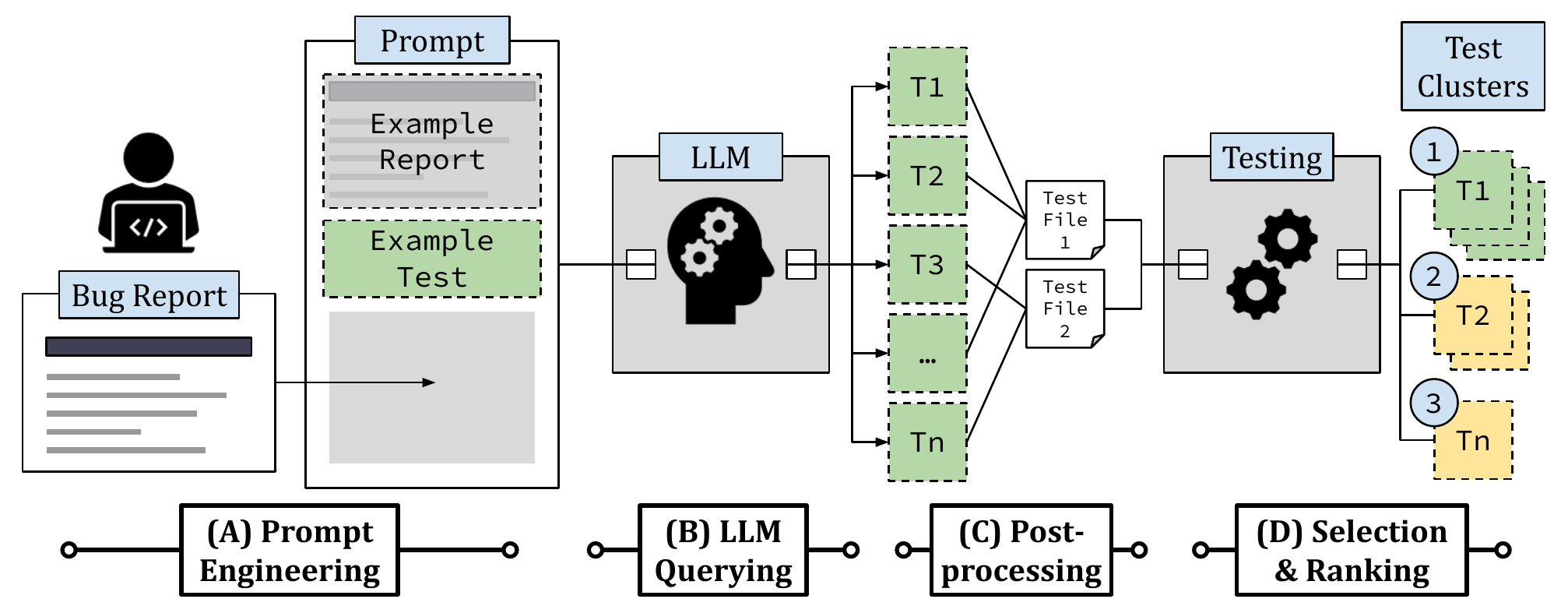}
\caption{Overview of \name\label{fig:overview}}
\end{figure}

An overview diagram of our approach is presented in \Cref{fig:overview}. Given 
a bug report, \name first constructs a prompt to query an LLM 
(\Cref{fig:overview}:(A)). Using this prompt, an initial set of test candidates 
are generated by querying the LLM multiple times (\Cref{fig:overview}:(B)).
Then, \name processes the tests to make them executable in the target program 
(\Cref{fig:overview}:(C)). \name subsequently identifies and curates tests that 
are likely to be bug reproducing, and if so, ranks them to minimize developer 
inspection effort (\Cref{fig:overview}:(D)). The rest of this section explains each stage in more detail using the running example provided in \Cref{tab:ex-report-math-63}.

\subsection{Prompt Engineering}

LLMs are, at the core, large autocomplete neural networks: prior work have found that different ways of `asking' the LLM to solve a problem will lead
to significantly varying levels of performance~\cite{Kojima2022Zreason}. Finding the best query to accomplish the given task is 
known as \textit{prompt engineering}~\cite{reynolds2021prompt}.

To make an LLM to generate a test method from a given bug report, we construct 
a Markdown document, which is to be used in the prompt, from the bug report: 
consider the example in Listing~\ref{lst:ex-prompt-math-63}, which is a 
Markdown document constructed from the bug report shown in 
\Cref{tab:ex-report-math-63}. \name adds a few distinctive parts to the Markdown document: the command ``Provide a self-contained example that 
reproduces this issue'', the start of a block of code in Markdown, (i.e., 
\texttt{\textasciigrave\textasciigrave\textasciigrave}), and finally the partial code snippet \code{public void test} 
whose role is to induce the LLM to write a test method.
\begin{table}[h!]
    \caption{\label{tab:ex-report-math-63}Example bug report (Defects4J Math-63).}
    \begin{tabular}{@{}lp{0.4\textwidth}@{}}
    \toprule
    Issue No.   & MATH-370\tablefootnote{\url{https://issues.apache.org/jira/browse/MATH-370}}\\ \midrule
    Title     & NaN in ``equals'' methods\\ \midrule
    Description & \begin{tabular}[c]{@{}l@{}}In ``MathUtils'', some ``equals'' methods will return true if \\ 
    both argument are NaN. Unless I'm mistaken, this contradicts \\ the IEEE standard.\\ 
    If nobody objects, I'm going to make the changes.\end{tabular} \\ \bottomrule
    \end{tabular}
\end{table}

\begin{lstlisting}[basicstyle=\footnotesize\ttfamily,
    columns=flexible,
    breaklines=true,
    caption={Example prompt without examples.},
    label={lst:ex-prompt-math-63},]
# NaN in "equals" methods
## Description
In "MathUtils", some "equals" methods will return true if both argument are NaN.  
Unless I'm mistaken, this contradicts the IEEE standard.
If nobody objects, I'm going to make the changes.

## Reproduction
>Provide a self-contained example that reproduces this issue.
```
public void test
\end{lstlisting}

We evaluate a range of variations of this basic prompt. Brown et al.~\cite{
brown2020language} report that LLMs benefit from question-answer examples 
provided in the prompt. In our case, this means providing examples of bug 
reports (questions) and the corresponding bug reproducing tests (answers). With 
this in mind, we experiment with a varying number of examples, to see whether 
adding more examples, and whether having examples from within the same project 
or from other projects, significantly influences performance.

As there is no real restriction to the prompt format, we also experiment with 
providing stack traces for crash bugs (to simulate situations where a stack 
trace was provided), or providing constructors of the class where the fault is 
located (to simulate situations where the location of the bug is reported).

Our specific template format makes it highly unlikely that prompts we generate
exist verbatim within the LLM training data. Further, most reports in practice
are only connected to the bug-revealing test via a chain of references. As 
such, our format partly mitigates data leakage concerns, among other steps 
taken to limit this threat described later in the manuscript.

\subsection{Querying an LLM}

Using the generated prompt, \name queries the LLM to predict the tokens that 
would follow the prompt. Due to the nature of the prompt, it is likely to 
generate a test method, especially as our prompt ends with the sequence 
\code{public void test}. We ensure that the result only spans the test method 
by accepting tokens until the first occurrence of the string \code{```}, which 
indicates the end of the code block in Markdown.

It is known that LLMs yield inferior results when performing completely greedy decoding (i.e., decoding strictly based on the most likely next token)~\cite{brown2020language}: they perform better when they are doing weighted random 
sampling, a behavior modulated by the \textit{temperature} parameter. Following 
prior work, we set our temperature to 0.7~\cite{brown2020language}, which allows
the LLM to generate multiple distinct tests based on the exact same prompt. We take the 
approach of generating multiple candidate reproducing tests, then using their 
characteristics to identify how likely it is that the bug is actually 
reproduced.

An example output from the LLM given the prompt in 
Listing~\ref{lst:ex-prompt-math-63} is shown in 
Listing~\ref{lst:ex-test-success-math-63}: at this point, the outputs from the LLM typically cannot be compiled on their own, and need other constructs such as import statements. We next present how \name integrates a generated test into the existing test suite to make it executable.

\begin{lstlisting}[basicstyle=\footnotesize\ttfamily,
    columns=flexible,
    breaklines=true,
    language=java,
    caption={Example LLM result from the bug report described in Table~\ref{tab:ex-report-math-63}.},
    label={lst:ex-test-success-math-63},]
public void testEquals() {
    assertFalse(MathUtils.equals(Double.NaN, Double.NaN));
    assertFalse(MathUtils.equals(Float.NaN, Float.NaN));
}
\end{lstlisting}

\subsection{Test Postprocessing}

We first describe how \name injects a test method into an existing suite then how \name resolves the remaining unmet dependencies.

\subsubsection{Injecting a test into a suitable test class}
If a developer finds a test method in a bug report, they will likely insert it 
into a test class which will provide the required context for the test method 
(such as the required dependencies). For example, for the bug in our running 
example, the developers added a reproducing test to the \code{MathUtilsTest} class, where 
most of the required dependencies are already imported, including the focal 
class, \code{MathUtils}. Thus, it is natural to also inject LLM-generated tests 
into existing test classes, as this matches developer workflow, while resolving 
a significant number of initially unmet dependencies.

\begin{lstlisting}[basicstyle=\footnotesize\ttfamily,
    columns=flexible,
    breaklines=true,
    language=java,
    caption={Target test class to which the test in Listing~\ref{lst:ex-test-success-math-63} is injected.},
    label={lst:ex-injected-class-math-63},]
public final class MathUtilsTest extends TestCase {
    ...
    public void testArrayEquals() {
        assertFalse(MathUtils.equals(new double[] { 1d }, null));
        assertTrue(MathUtils.equals(new double[] {
            Double.NaN, Double.POSITIVE_INFINITY,
        ...
\end{lstlisting}

To find the best test class to inject our test methods into, we find the test class
that is \emph{lexically} most similar to the generated test 
(Algorithm~\ref{alg:1-test-postprocessing}, line 1). The intuition is that, 
if a test method belongs to a test class, the test method likely uses similar 
methods and classes, and is thus lexically related, to other tests from that 
test class. Formally, we assign a matching score for each test class based on \Cref{eq:test2class}:
\begin{equation}
    sim_{c_i} = |T_{t} \cap T_{c_i}| / |T_{t}| 
    \label{eq:test2class}
\end{equation}
where $T_t$ and $T_{c_i}$ are the set of tokens in the generated test method 
and the $i$th test class, respectively. 
As an example, Listing~\ref{lst:ex-injected-class-math-63} shows the 
key statements of the \code{MathUtilsTest} class. Here, the test class contains 
similar method invocations and constants with those used by the LLM-generated 
test in Listing~\ref{lst:ex-test-success-math-63}, particularly in
lines 4 and 6.

As a sanity check, we inject ground-truth developer-added bug reproducing tests 
from the Math and Lang projects of the Defects4J benchmark, and check if they 
execute normally based on Algorithm~\ref{alg:1-test-postprocessing}. We find 
execution proceeds as usual for 89\% of the time, suggesting that the algorithm 
reasonably finds environments in which tests can be executed.

\RestyleAlgo{ruled} 
\SetKwComment{Comment}{/* }{ */}
\begin{algorithm}[t]
\footnotesize
\caption{Test Postprocessing}\label{alg:1-test-postprocessing}
\KwIn{A test method $tm$; Test suite $\mathcal{T}$ of SUT; source code files $\mathcal{S}$ of SUT;}
\KwOut{Updated test suite $\mathcal{T}'$}% with an injected test method and dependencies;}

$c_{best} \leftarrow$ \codealgo{findBestMatchingClass\(}$tm$, $\mathcal{T}$\codealgo{\)}\;
$deps \leftarrow$ \codealgo{getDependencies\(}$tm$\codealgo{\)}\;
$needed\_deps \leftarrow$ \codealgo{getUnresolved\(}$deps$, $c_{best}$\codealgo{\)}\;
$new\_imports \leftarrow$ \codealgo{set()}\;    
\For{$dep$ in $needed\_deps$}{
    $target \leftarrow$ \codealgo{findClassDef\(}$dep$, $\mathcal{S}$\codealgo{\)}\;
    \eIf{$target$ is \codealgo{null}}{
        $new\_imports$\codealgo{.add\(findMostCommonImport\(}$dep, \mathcal{S}, \mathcal{T}$\codealgo{\)\)}\;
    }{
        $new\_imports$\codealgo{.add\(}$target$\codealgo{\)}\;
    }
}
$\mathcal{T}' \leftarrow$ \codealgo{injectTest\(}$tm$, $c_{best}$, $\mathcal{T}$\codealgo{\)}\;
$\mathcal{T}' \leftarrow$ \codealgo{injectDependencies\(}$new\_imports, c_{best}, \mathcal{T}'$\codealgo{\)}\;
\end{algorithm}

\subsubsection{Resolving remaining dependencies}
Although many dependency issues are resolved by placing the test in the right 
class, the test may introduce new constructs that need to be imported.
To handle these cases, \name heuristically infers packages to import. 

Line $2$ to $10$ in Algorithm~\ref{alg:1-test-postprocessing} describe the 
dependency resolving process of \name. First, \name parses the generated test 
method and identifies variable types and referenced class 
names/constructors/exceptions. \name then filters ``already imported'' class 
names by lexically matching names to existing import statements in the test 
class (Line $3$). 

As a result of this process, we find types that are not resolved within the
test class. \name first attempts to find public classes with the identified  
name of the type; if there is exactly one such file, the classpath to the 
identified class is derived (Line $7$), and an import statement is added 
(Line $11$). However, either no or multiple matching classes may exist. In both 
cases, \name looks for \code{import} statements ending with the target class 
name within the project (e.g., when searching for \code{MathUtils}, \name looks 
for \code{import .*MathUtils;}). \name selects the most common import statement 
across all project source code files. Additionally, we add a few rules that 
allow assertion statements to be properly imported, even when there are no 
appropriate imports within the project itself.

Our postprocessing pipeline does not guarantee compilation in all cases, but 
the heuristics used by \name are capable of resolving most of the unhandled 
dependencies of a raw test method. After going through the postprocessing 
steps, \name executes the tests to identify candidate bug reproducing 
tests. 

\subsection{Selection and Ranking}
\label{sec:sel_n_rank}

A test is a Bug Reproducing Test (BRT) if and only if the test fails due to the 
bug specified in the report. A \emph{necessary} condition for a test generated 
by \name to be a BRT is that the test compiles and fails in the buggy program: we call such tests FIB (Fail In the Buggy program) tests. 
However, not all FIB tests are BRTs, making it difficult to tell whether bug 
reproduction has succeeded or not. This is one factor that separates us from 
crash reproduction work~\cite{Soltani2020aa}, as crash reproduction techniques can confirm whether the bug has been reproduced by comparing the stack traces at the time of crash. 
On the other hand, it is imprudent to present all generated FIB tests 
to developers, as asking developers to iterate over multiple solutions is 
generally undesirable~\cite{Kochhar2016FLSurvey, Noller2022APRSurvey}.
As such, \name attempts to decide when to suggest a test and, if so, which test 
to suggest, using several patterns we observe to be correlated to successful 
bug reproductions. 

\SetKwComment{Comment}{/* }{ */}
\SetKwProg{Fn}{Function}{ is}{end}
\begin{algorithm}[t]
    \footnotesize
\caption{Test Selection and Ranking}\label{alg:2-test-ranking}
\KwIn{Pairs of modified test suites and injected test methods $\mathcal{S_{T'}}$; target program with bug $P_{b}$;  bug report $BR$; agreement threshold $Thr$;}
\KwOut{Ordered list of ranked tests $ranking$; }
$FIB \leftarrow $ \codealgo{set()}\; 
\For{$(\mathcal{T}', tm_i) \in \mathcal{S_{T'}}$}{
    $r \leftarrow$ \codealgo{executeTest(}$\mathcal{T'}, P_{b}$\codealgo{)}\;
    \If{\codealgo{hasNoCompileError(}$r$\codealgo{) && isFailed(}$tm_i, r$\codealgo{)}}{
        $FIB$\codealgo{.add(}$(tm_i, r)$\codealgo{)}\;
    }
}

$clusters\leftarrow$\codealgo{clusterByFailureOutputs(}$FIB$\codealgo{)}\;
$output\_clus\_size \leftarrow clusters$\codealgo{.map(size)}\;
$max\_output\_clus\_size \leftarrow $\codealgo{max(}$output\_clus\_size$\codealgo{)}\;
\If{$max\_output\_clus\_size \leq Thr$}{
    \Return{\codealgo{list()}}\;
}
$FIB_{uniq} \leftarrow$ \codealgo{removeSyntacticEquivalents(}$FIB$\codealgo{)}\;

$br\_output\_match \leftarrow clusters$\codealgo{.map(matchOutputWithReport(}$BR$\codealgo{))}\;
$br\_test\_match \leftarrow FIB_{uniq}$\codealgo{.map(matchTestWithReport(}$BR$\codealgo{))}\;
$tok\_cnts \leftarrow FIB_{uniq}$\codealgo{.map(countTokens)}\;
$ranking \leftarrow $\codealgo{list()}\; 
$clusters \leftarrow clusters$\codealgo{.sortBy(}$br\_output\_match, output\_clus\_size,tok\_cnts$\codealgo{)}\;
\For{$clus \in clusters$}{
    $clus \leftarrow clus$\codealgo{.sortBy(}$br\_test\_match, tok\_cnts$\codealgo{)}\;
}
\For{$i=0; i<$\codealgo{max(}$output\_clus\_size$\codealgo{)}$;i\gets i+1$}{
\For{$clus \in clusters$}{
    \If{$i<clus$\codealgo{.length()}}{$ranking$\codealgo{.push(}$clus$\codealgo{[i])}}
}
}
\Return{$ranking$}\;

\end{algorithm}

Algorithm~\ref{alg:2-test-ranking} outlines how \name decides whether to 
present results and, if so, how to rank the generated tests. In Line 1-10, \name 
first decides whether to show the developer any results at all (selection). We 
group the FIB tests that exhibit the same failure output (the same error type 
and error message) and look at the number of the tests in the same group (the 
$max\_output\_clus\_size$ in Line 8). This is based on the intuition that, if 
multiple tests show similar failure behavior, then it is likely that the LLM is 
`confident' in its predictions as its independent predictions `agree' with each other, and 
there is a good chance that bug reproduction has succeeded. \name can be 
configured to only show results when there is significant agreement in the 
output (setting the agreement threshold $Thr$ high) or show more exploratory 
results (setting $Thr$ low).

Once it decides to show its results, \name relies on three heuristics to rank 
generated tests, in the order of increasing discriminative strength.
First, tests are likely to be bug reproducing if the fail message and/or the 
test code shows the behavior (exceptions or output values) observed and 
mentioned in the bug report. While this heuristic is precise, its decisions are not very discriminative, as tests can only be divided into groups of 
`contained' versus `not contained'. Next, we look at the `agreement' between 
generated tests by looking at output cluster size ($output\_clus\_size$), which 
represents the `consensus' of the LLM generations. Finally, \name prioritizes 
based on test length (as shorter tests are easier to understand), which is the 
finest-grained signal. We first leave only syntactically unique tests 
(Line 11), then sort output clusters and tests within those clusters 
using the heuristics above (Lines 16 and 18). 

As tests with the same failure output are similar to each other, we expect 
that, if one test from a cluster is not BRT, the rest from the same cluster are 
likely not BRT as well. Hence, \name shows tests from a diverse array of 
clusters. For each $i$th iteration in Line 19-22, the $i$th ranked 
test from each cluster is selected and added to the list.

\section{Evaluation}
\label{sec:evaluation}
This section provides evaluation details for our experiments.

\subsection{Dataset} 
As a comprehensive evaluation benchmark, we use \textit{Defects4J} version 2.0, 
which is a manually curated dataset of real-world bugs gathered from 17 Java 
projects. Each Defects4J bug is paired to a corresponding bug report\footnote{
Except for the Chart project, for which only 8 bugs have reports}, which makes 
the dataset ideal for evaluating the performance of \name. Among 814 bugs that 
have a paired bug report, we find that 58 bugs have incorrect pairings, while six bugs have different directory structures between the buggy and fixed 
versions: this leaves \textbf{750} bugs for us to evaluate \name on. 60 bugs in 
the Defects4J benchmark are included in the JCrashPack~\cite{soltani2020benchmark} dataset used in the 
crash reproduction literature; we use this subset when comparing against 
crash reproduction techniques.

As Codex, the LLM we use,
was trained with data collected until July 2021, the Defects4J dataset 
is not free from data leakage concerns, even if the prompt format we use is 
unlikely to have appeared verbatim in the data. To mitigate such concerns, from 
17 GitHub repositories\footnote{These repositories have been manually chosen 
from either Defects4J projects that are on GitHub and open to new issues, or Java projects that have been modified since 10th July 2022 with at least 100 or more stars, as of 1st of August 2022. A list of 17 repositories is available in our artifact.} that use JUnit, we gather 581 Pull Requests (PR) created after the Codex training data cutoff point, ensuring that 
this dataset could not have been used to train Codex. We further 
check if a PR adds any test to the project (435 left after discarding  
non-test-introducing ones), and filter out the PRs that are not merged to the 
main branch or associated with multiple issues (84 left). 

For these 84 PRs, we verify that the bug can be reproduced by checking that a 
developer-added test added by the PR fails on the pre-merge commit without 
compilation errors, and passes on the post-merge commit. We add the pair to our 
final list only when all of them have been reproduced. After the final check, 
we end up with \textbf{31} reproducible bugs and their bug reports. This 
dataset is henceforth referred to as the GHRB (GitHub Recent Bugs) dataset. We 
use this dataset to verify that trends observed in Defects4J are not due to 
data leakage.

\subsection{Metrics} 
A test is treated as a Bug Reproducing Test (BRT) in our evaluation if 
it fails on the version that contains the bug specified in the report, and 
passes on the version that fixes the bug. We say that a bug is \textit{reproduced} if 
\name generates at least one BRT for that bug. The number of bugs that are 
reproduced is counted for each evaluation technique.

We use the \texttt{PRE\_FIX\_REVISION} and \texttt{POST\_FIX\_REVISION} 
versions in the Defects4J benchmarks as the buggy/fixed versions, respectively.
The two versions reflect the \textit{actual} state of the project when the bug 
was discovered/fixed. For the GHRB dataset, as we gathered the data based on 
code changes from merged pull requests, we use pre-merge and post-merge 
versions as the buggy/fixed versions.

EvoCrash~\cite{Soltani2020aa}, the crash reproduction technique we compare with,
originally checks whether the \textit{crash stack} is reproduced in the \emph{buggy 
version}. For fair comparison, we evaluate EvoCrash under our reproduction
criterion: EvoCrash-generated tests are executed on the buggy and fixed versions,
and when execution results change, we treat the test as a BRT.

To evaluate the rankings produced by \name, we focus on two aspects: 
the capability of \name to rank the actual BRTs higher, and the degree of 
effort required from developers to inspect the ranked tests. For the former, we 
use the $acc@n$ metric, which counts the number of bugs whose BRTs are found 
within the top $n$ places in the ranking. Additionally, we report the precision 
of \name by dividing $acc@n$ with the number of all selected bugs, representing 
how often a developer would accept a suggestion by \name. To estimate developer effort, we 
use the $wef$ metric: the number of non-reproducing tests ranked 
higher than any bug reproducing test. If there are no BRTs, we 
report $wef$ as the total number of the target FIB tests in ranking. 
We also use $wef@n$, which shows the wasted effort when using 
the top $n$ candidates. 

\subsection{Environment} 

All experiments are performed on a machine running Ubuntu 18.04.6 LTS, 
with 32GB of ram and Intel(R) Core(TM) i7-7700 CPU @ 3.60GHz CPU. We access 
OpenAI Codex via its closed beta API, using the \code{code-davinci-002} model. For Codex, we set the temperature to 0.7, and the maximum 
number of tokens to 256. We script our experiments using Python 3.9, and parse 
Java files with the \texttt{javalang} library~\cite{c2nes2022javalang}.
Our replication package is online\footnote{\url{https://anonymous.4open.science/r/llm-testgen-artifact-2753}}.

\section{Research Questions}
\label{sec:rqs}
We aim to answer the following research questions. 
\subsection{RQ1: Efficacy}

With RQ1, we seek to quantitatively evaluate the performance of
\name using the Defects4J benchmark.

\begin{itemize}
    \item \textbf{RQ1-1: How many bug reproducing tests can \name generate?} We evaluate how many bugs in total are reproduced by \name using various prompt settings.
    \item \textbf{RQ1-2: How does \name compare to other techniques?} In the absence of generic report-to-test techniques, we compare against EvoCrash, a crash reproduction technique. We also compare against a `Copy\&Paste' baseline that directly uses code snippets (identified with the HTML \code{<pre>} tag or via infoZilla~\cite{Premraj2008infoZilla}) within the bug report as tests. For code that could be parsed as a Java compilation unit, we add the code as a test class and add JUnit imports if necessary to run it as a test. Otherwise, we wrap the code snippet in a test method and evaluate it under the same conditions as \name.
\end{itemize}

\subsection{RQ2: Efficiency} % Efficiency?

With RQ2, we examine the efficiency of \name in terms of the amount of
resources it uses, to provide an estimate of the costs of deploying
\name in a real-world context.

\begin{itemize}
    \item \textbf{RQ2-1: How many Codex queries are required?} We estimate how many queries are needed to achieve a certain bug-reproduction rate on the Defects4J dataset based on a pool of generated tests.
    \item \textbf{RQ2-2: How much time does \name need?} Our technique consists of querying an LLM, making it executable, and ranking: we measure the time taken at each stage.
    \item \textbf{RQ2-3: How many tests should the developer inspect?} We evaluate how many bugs could be reproduced within 1, 3, and 5 suggestions, along with the amount of `wasted effort' required from the developer.
\end{itemize}

\subsection{RQ3: Practicality}

Finally, with RQ3, we aim to investigate how well \name generalizes by applying 
it to the GHRB dataset. 

\begin{itemize}
    \item \textbf{RQ3-1: How often can \name reproduce bugs in the wild?} To mitigate data leakage issues, we evaluate \name on the GHRB dataset, checking how many bugs can be reproduced on it.
    \item \textbf{RQ3-2: How reliable are the selection and ranking techniques of \name?} 
    We investigate whether the factors that were used during selecting bugs and ranking tests for the Defects4J dataset are still valid for the GHRB dataset, and thus can be used for other projects in general.
    \item \textbf{RQ3-3: What does reproduction success and failure look like?}
    To provide qualitative context to our results, we describe examples of bug reproduction success and failure from the GHRB dataset.
\end{itemize}

\section{Experimental Results}
\label{sec:results}

\subsection{RQ1. How effective is \name?}

\subsubsection{RQ1-1}
\Cref{tab:d4j_per_prompt} shows which prompt/information settings work best, where
$n=N$ means we queried the LLM $N$ times for reproducing tests.
When using examples from the source project, we use the oldest tests available 
within that project; otherwise, we use two handpicked report-test pairs 
(Time-24, Lang-1) throughout all projects. We find that providing constructors 
(\textit{\`{a} la} AthenaTest~\cite{tufano2020unit}) does not help 
significantly, but adding stack traces does help reproduce crash bugs, 
indicating that \name can benefit from using the stack information to replicate 
issues more accurately. Interestingly, adding within-project examples shows 
poorer performance: inspection of these cases has revealed that, in such cases, 
\name simply copied the provided example even when it should not have, leading 
to lower performance. We also find that the number of examples makes a 
significant difference (two-example n=10 values are sampled from n=50 results 
from the default setting), confirming the existing finding that adding examples helps improve
performance. In 
turn, the number of examples seems to matter less than the number of times the 
LLM is queried, as we further explore in RQ2-1. As the two-example n=50 
setting shows the best performance, we use it as the default setting 
throughout the rest of the paper.

\begin{table}[ht]
    \centering
    \caption{Reproduction performance for different prompts\label{tab:d4j_per_prompt}}
    \scalebox{0.8}{
    \begin{tabular}{lcc}
    \toprule
    Setting & reproduced & FIB \\\midrule
    No Example (n=10) & 124 & 440 \\
    One Example (n=10) & 166 & 417 \\
    One Example from Source Project (n=10) & 152 & 455 \\
    One Example with Constructor Info (n=10) & 167 & 430 \\
    Two Examples (n=10, 5th percentile) & 161 & 386 \\
    Two Examples (n=10, median) & 173 & 409 \\
    Two Examples (n=10, 95th percentile) & 184 & 429 \\
    Two Examples (n=50) & \textbf{251} & \textbf{570} \\\midrule
    One Example, Crash Bugs (n=10) & 69 & 153 \\
    One Example with Stack, Crash Bugs (n=10) & 84 & 155 \\
    \bottomrule
    \end{tabular}
    }
\end{table} 

Under the two-example n=50 setting, we find that overall \textbf{251} 
bugs, or 33.5\% of 750 studied Defects4J bugs, are reproduced by \name. 
\Cref{tab:d4j_per_proj} presents a breakdown of the performance per project. 
While there is at least one bug reproduced for every project, the proportion
of bugs reproduced can vary significantly. For example, \name reproduces a 
small number of bugs in the Closure project, which is known to have a unique 
test structure~\cite{Martinez2016AutomaticRO}. On the other hand, the 
performance is stronger for the Lang or Jsoup projects, whose tests are 
generally self-contained and simple. Additionally, we find that the average
length of the generated test body is about 6.5 lines (excluding comments
and whitespace), indicating \name is capable of writing meaningfully long
tests.

\begin{table}[ht]
    \centering
    \caption{Bug reproduction per project in Defects4J:\protect\\
    x/y means x reproduced out of y bugs\label{tab:d4j_per_proj}}
    % Lang's 63 total bugs stems from Lang2(deprecated) and Lang30 (excluded by us).
    \scalebox{0.9}{
    \begin{tabular}{lr|lr|lr}
    \toprule
    Project & rep/total & Project & rep/total & Project & rep/total\\\midrule
    Chart & 5/7 & Csv & 6/16 & JxPath & 3/19 \\
    Cli & 14/29 & Gson & 7/11 & Lang & 46/63 \\
    Closure & 2/172 & JacksonCore & 8/24 & Math & 43/104 \\
    Codec & 10/18 & JacksonDatabind & 30/107 & Mockito & 1/13 \\
    Collections & 1/4 & JacksonXml & 2/6 & Time & 13/19 \\
    Compress & 4/46 & Jsoup & 56/92 & \textbf{Total} & \textbf{251/750} \\
    \bottomrule
    \end{tabular}
    }
\end{table} 

\begin{tcolorbox}[boxrule=0pt,frame hidden,sharp corners,enhanced,borderline north={1pt}{0pt}{black},borderline south={1pt}{0pt}{black},boxsep=2pt,left=2pt,right=2pt,top=2.5pt,bottom=2pt] % to completely remove left/right rules
    \textbf{Answer to RQ1-1:} A large (251) number of bugs can be replicated
    automatically, with bugs replicated over a diverse group of projects.
    Further, the number of examples in the prompt and the number of generation 
    attempts have a strong effect on performance.
\end{tcolorbox}  

\subsubsection{RQ1-2} 
\begin{figure}[h!]
    \centering
    \begin{subfigure}{0.15\textwidth}
        \centering
        \includegraphics[width=\linewidth]{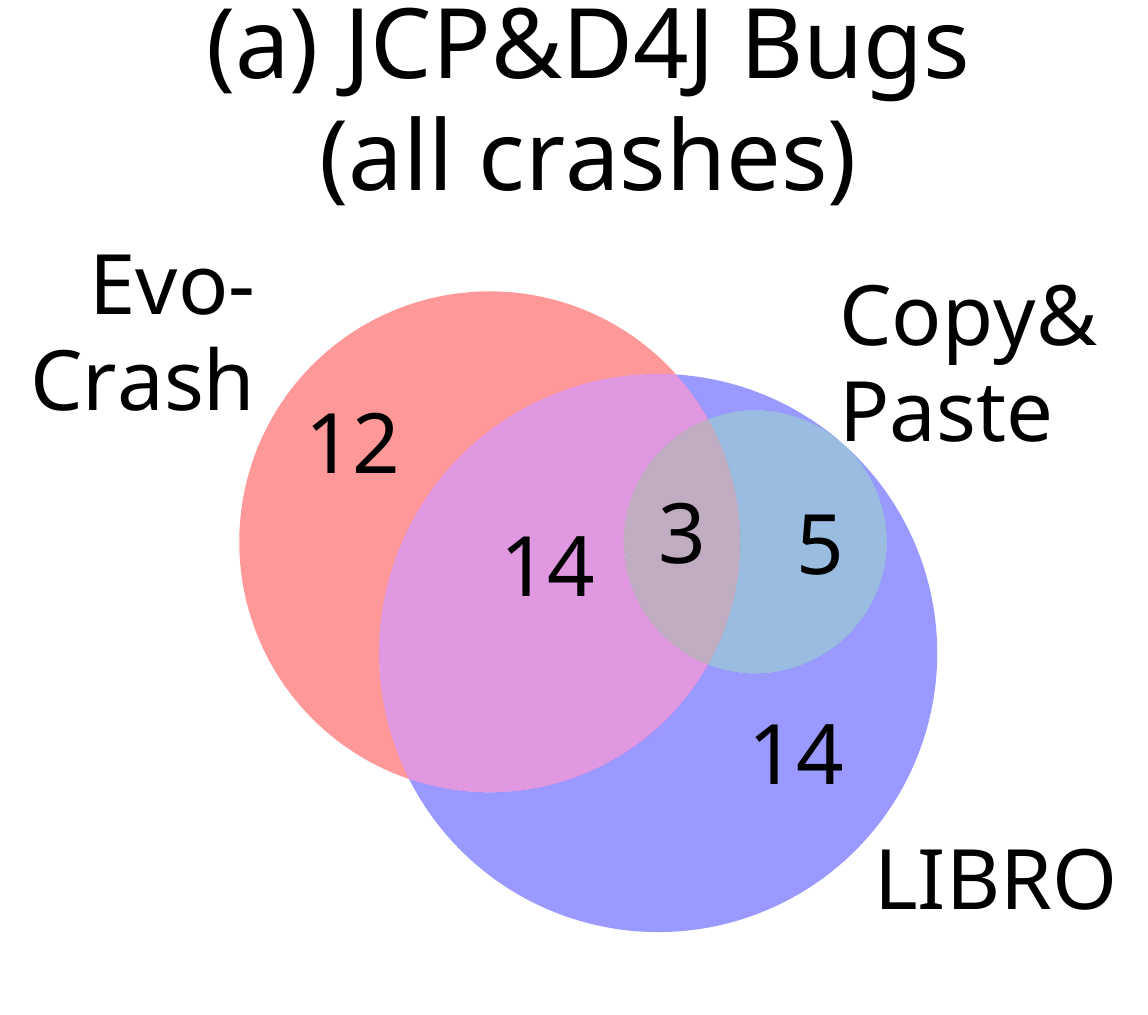}
    \end{subfigure}
    \begin{subfigure}{0.135\textwidth}
        \centering
        \includegraphics[width=\linewidth]{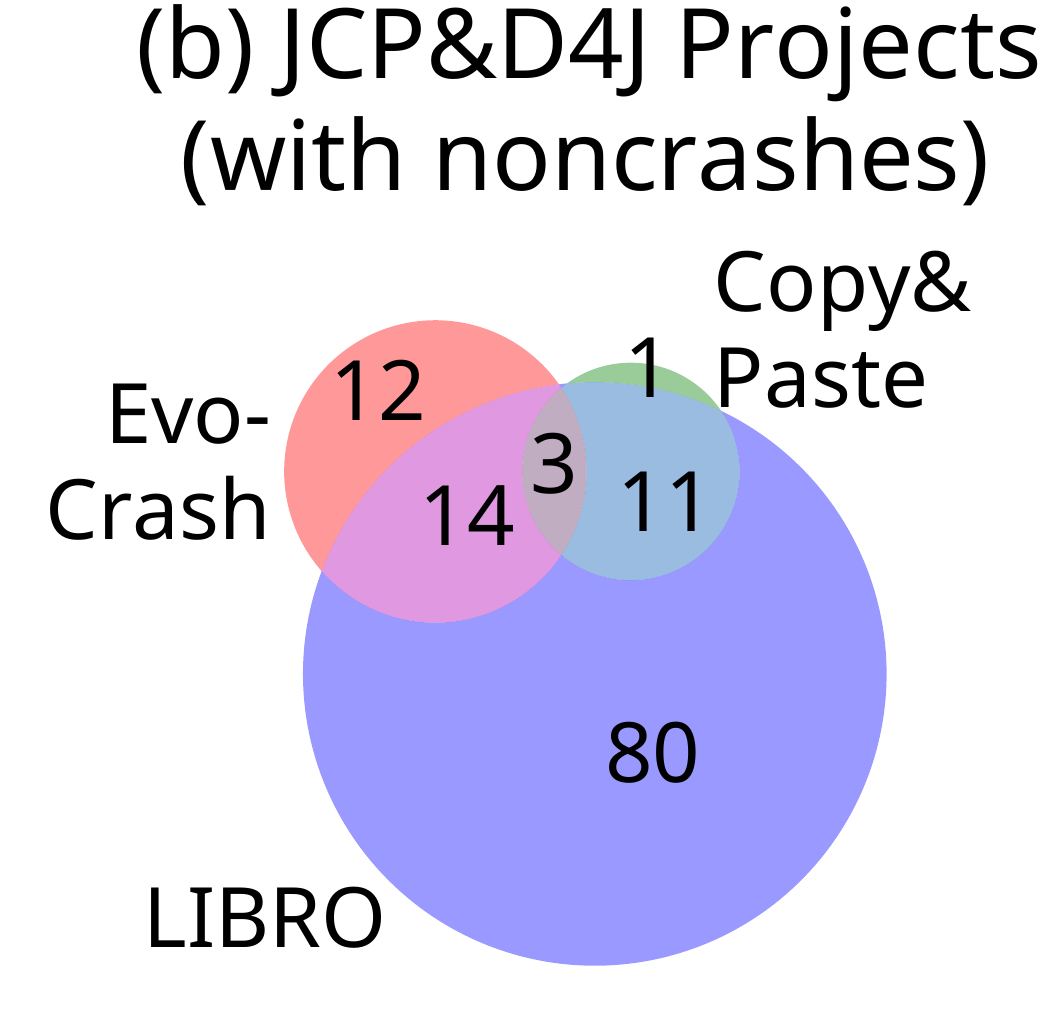}
    \end{subfigure}
    \begin{subfigure}{0.15\textwidth}
        \centering
        \includegraphics[width=\linewidth]{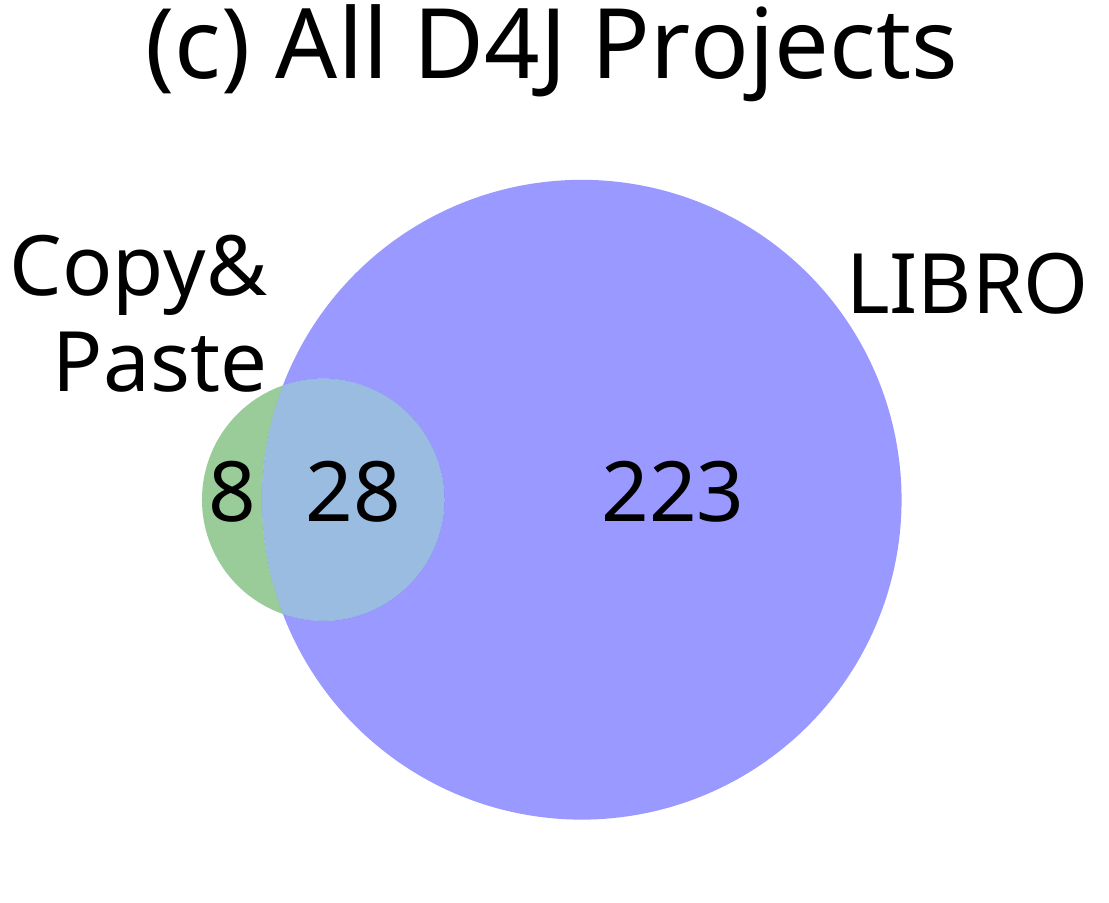}
    \end{subfigure}
    \caption{Baseline comparison on bug reproduction capability}
    \label{fig:baseline-venn}
  \end{figure}

We further compare \name against the 
state-of-the-art crash reproduction technique, EvoCrash, and the `Copy\&Paste 
baseline' that uses code snippets from the bug reports. We present the comparison results in Figure~\ref{fig:baseline-venn}. We find \name 
replicates a large and distinct group of bugs compared to other baselines. 
\name reproduced 91 more unique bugs (19 being crash bugs) than EvoCrash, which 
demonstrates that \name can reproduce non-crash bugs prior work could not 
handle (Fig.~\ref{fig:baseline-venn}(b)). On the other hand, the Copy\&Paste 
baseline shows that, while the BRT is sometimes included in the bug report, the 
report-to-test task is not at all trivial. Interestingly, eight bugs reproduced
by the Copy\&Paste baseline were not reproduced by \name; we find that this is due 
to long tests that exceed the generation length of \name, or due to dependency on 
complex helper functions.

\begin{tcolorbox}[boxrule=0pt,frame hidden,sharp corners,enhanced,borderline north={1pt}{0pt}{black},borderline south={1pt}{0pt}{black},boxsep=2pt,left=2pt,right=2pt,top=2.5pt,bottom=2pt]
    \textbf{Answer to RQ1-2:} \name is capable of replicating a large and distinct group of bugs
    relative to prior work.
\end{tcolorbox}  

\subsection{RQ2. How efficient is \name?}

\begin{figure}[h!]
    \centering
    \begin{subfigure}{0.235\textwidth}
        \centering
        \includegraphics[width=\linewidth]{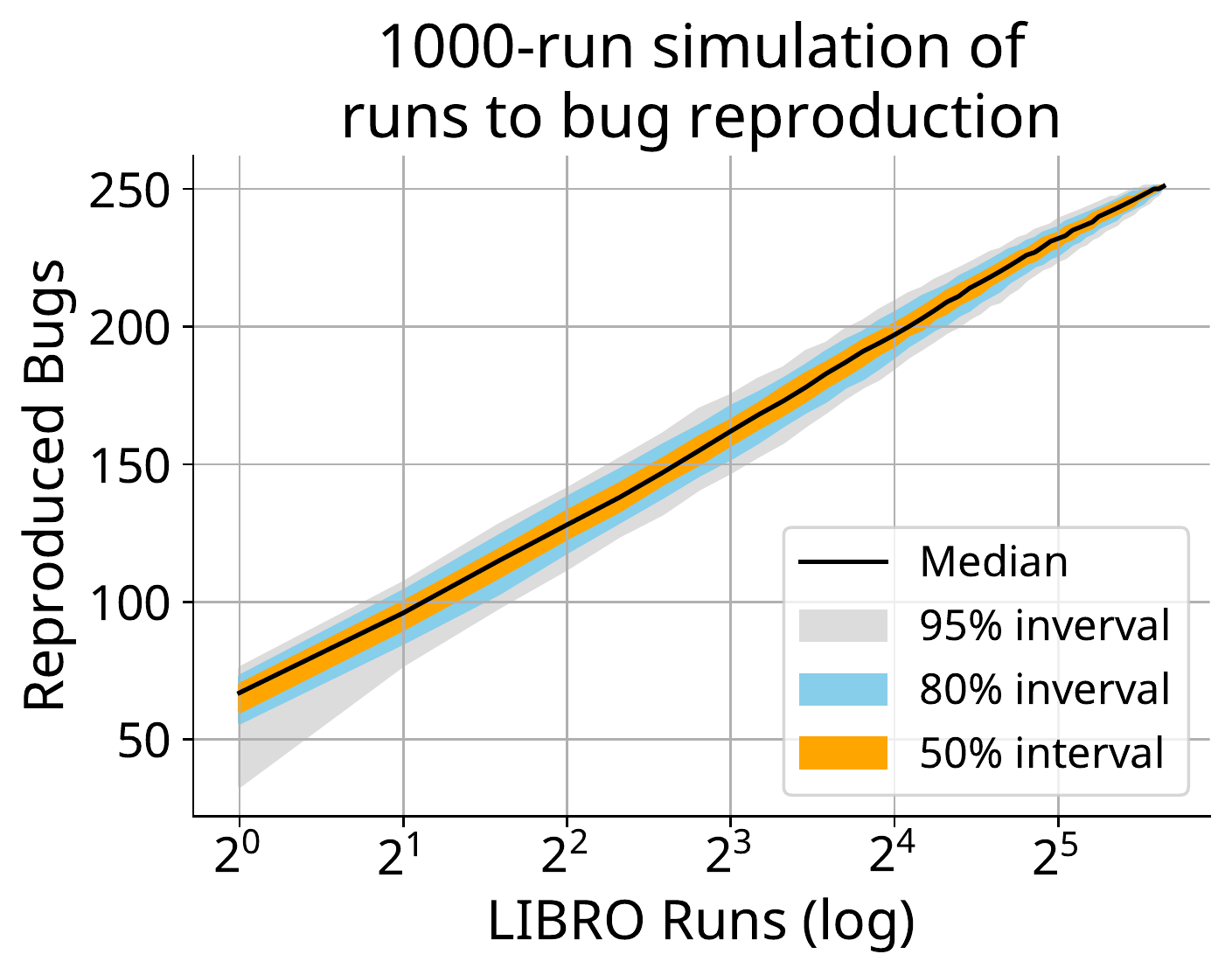}
    \end{subfigure}%
    \hfill
    \begin{subfigure}{0.235\textwidth}
        \centering
        \includegraphics[width=\linewidth]{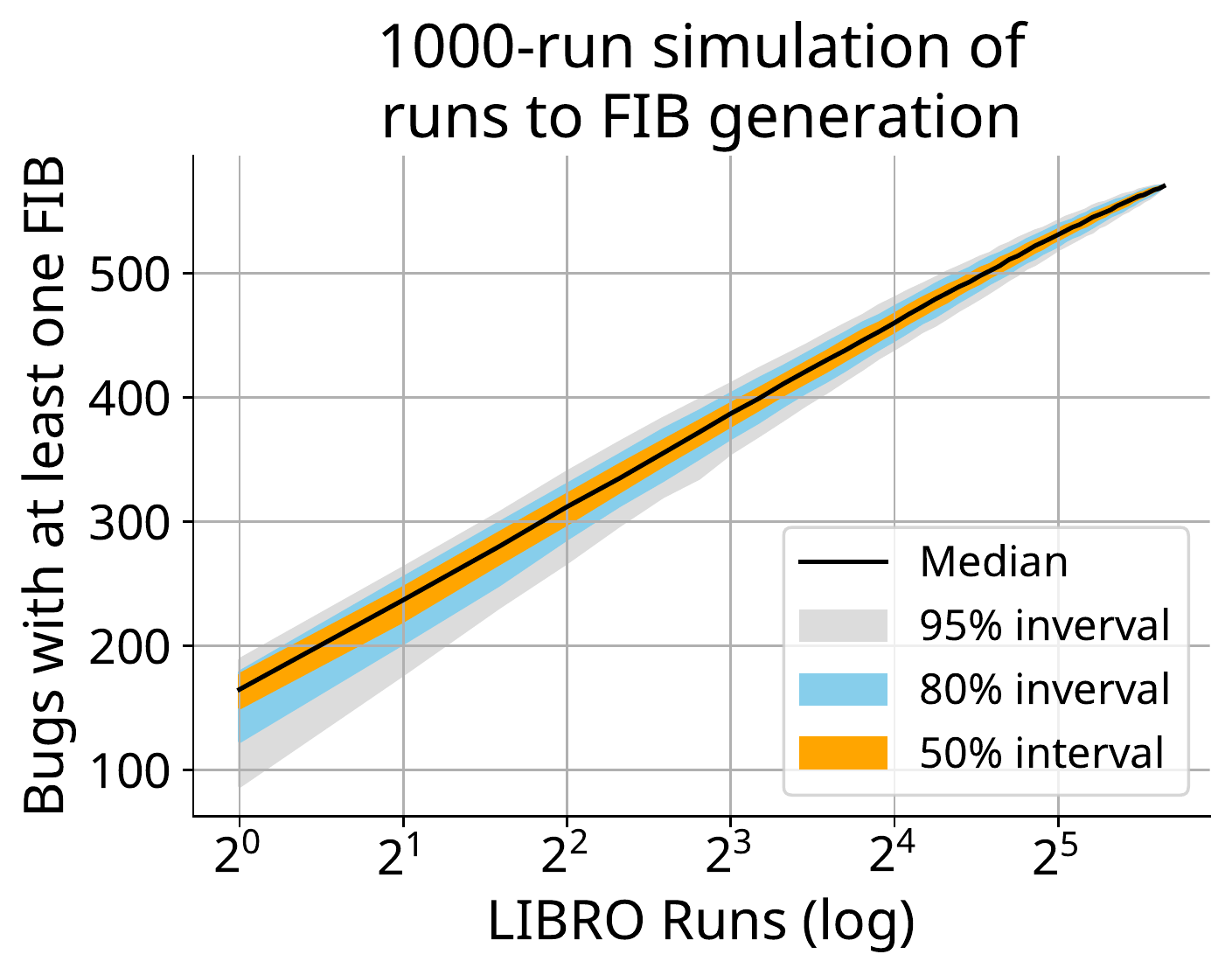}
    \end{subfigure}
    \caption{Generation attempts to performance. Left depicts bugs reproduced as attempts increase, right for FIB}
    \label{fig:run2perf}
  \end{figure}

\subsubsection{RQ2-1} Here, we investigate how many tests must be 
generated to attain a certain bug reproduction performance. To do so, for each 
Defects4J bug, we randomly sample $x$ tests from the 50 generated under the 
default setting, leaving a reduced number of tests per bug. We then check the 
number of bugs reproduced $y$ when using only those sampled tests. We repeat
this process 1,000 times to approximate the distribution.

The results are presented in \Cref{fig:run2perf}. Note that the $x$-axis is in
log scale. Interestingly, we find a logarithmic relation holds between the 
number of test generation attempts and the median bug reproduction performance. 
This suggests that it becomes increasingly difficult, yet stays possible, to 
replicate more bugs by simply generating more tests. As the graph shows no signs
of plateauing, experimenting with an even greater sample of tests may result in 
better bug reproduction results. 

\begin{tcolorbox}[boxrule=0pt,frame hidden,sharp corners,enhanced,borderline north={1pt}{0pt}{black},borderline south={1pt}{0pt}{black},boxsep=2pt,left=2pt,right=2pt,top=2.5pt,bottom=2pt]
    \textbf{Answer to RQ2-1:} The number of bugs reproduced increases
    logarithmically to the number of tests generated, with no sign of performance plateauing.
\end{tcolorbox}  

\begin{table}[ht]
    \centering
    \caption{The time required for the pipeline of \name\label{tab:time_performance}}
    \scalebox{0.9}{
    \begin{tabular}{ccccccc}
    \toprule
     & Prompt & API & Processing & Running & Ranking & Total\\ \midrule
    Single Run & <1 $\mu$s & 5.85s & 1.23s & 4.00s & - & 11.1s\\
    50-test Run & <1 $\mu$s & 292s & 34.8s & 117s & 0.02s & 444s\\
    \bottomrule
    \end{tabular}
    }
\end{table} 

\subsubsection{RQ2-2} We report the time it takes to perform each step of
our pipeline in \Cref{tab:time_performance}. We find API querying takes the 
greatest amount of time, requiring about 5.85 seconds. 
Postprocessing and test executions take 1.23 and 4 seconds per test (when the 
test executes), respectively. Overall, \name took an average of 444 seconds to generate 50 
tests and process them, which is well within the 10-minute search budget often 
used by search-based techniques~\cite{Soltani2020aa}.

\begin{tcolorbox}[boxrule=0pt,frame hidden,sharp corners,enhanced,borderline north={1pt}{0pt}{black},borderline south={1pt}{0pt}{black},boxsep=2pt,left=2pt,right=2pt,top=2.5pt,bottom=2pt]
    \textbf{Answer to RQ2-2:} Our time measurement suggests that \name does not take a significantly
    longer time than other methods to use.
\end{tcolorbox}

\subsubsection{RQ2-3}
\begin{figure}[ht]
    \centering
    \begin{subfigure}{0.232\textwidth}
        \centering
        \includegraphics[width=\linewidth]{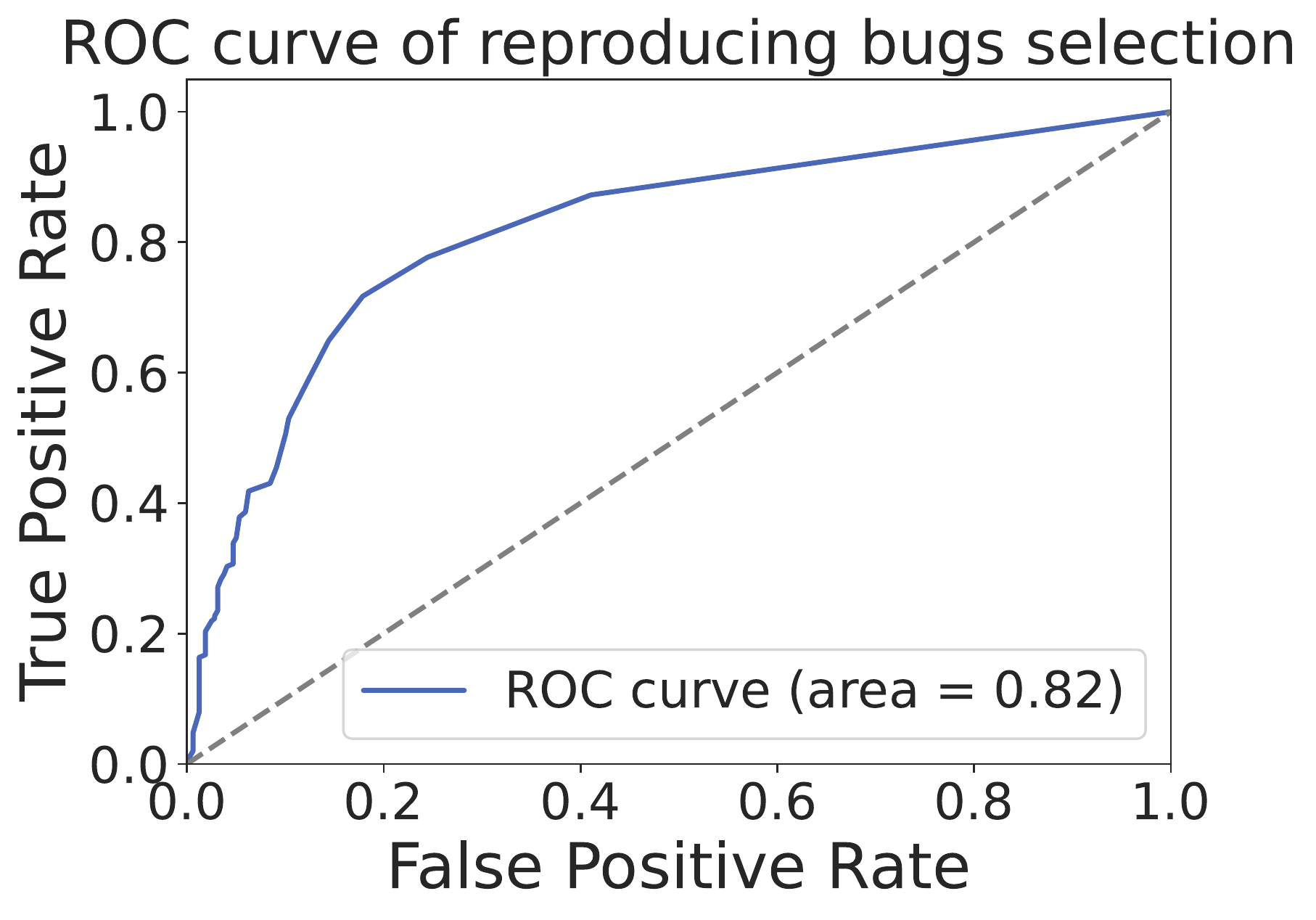}
    \end{subfigure}%
    \begin{subfigure}{0.243\textwidth}
        \centering
        \includegraphics[width=\linewidth]{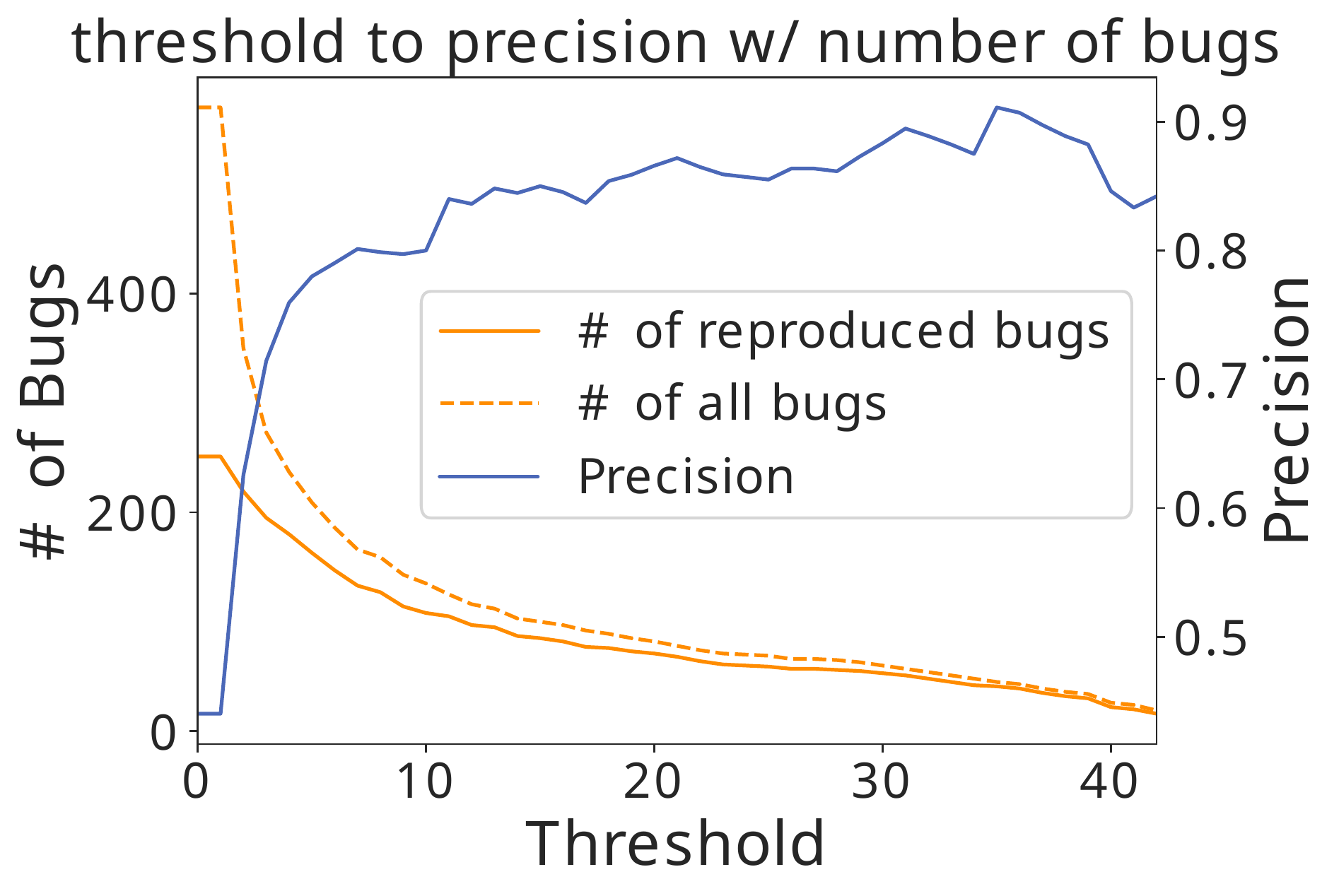}
    \end{subfigure}%
    \caption{ROC curve of bug selection (Left), Effect of thresholds to the number of bugs selected and precision (Right)}
    \label{fig:thr2precision}
\end{figure}

With this research question, we measure how effectively \name prioritizes bug reproducing tests 
via its selection and ranking procedure.
As \name only shows results above a certain agreement threshold, 
$Thr$ from \Cref{sec:sel_n_rank}, we first present the
trade-off between the number of total bugs reproduced and precision 
(i.e., the proportion of successfully reproduced bugs among all selected by \name)
in Figure~\ref{fig:thr2precision}. 
As we increase the threshold, more suggestions (including BRTs) are discarded,
but the precision gets higher, suggesting one can smoothly increase precision 
by tuning the selection threshold. 

We specifically set the agreement threshold to $1$, a conservative value, in order to preserve as many 
reproduced bugs as possible. Among the 570 bugs with a FIB, 350 bugs are selected. Of those 350, 219 are
reproduced (leading to a precision of $0.63 (=\frac{219}{350})$ 
whereas recall (i.e., proportion of selected reproduced bugs among all
reproduced bugs) is $0.87 (=\frac{219}{251})$. 
From the opposite perspective, the selection process 
filters out 188 bugs that were not reproduced, while dropping only a 
few successfully reproduced bugs. Note that if we set the threshold 
to $10$, a more aggressive value, we can achieve a higher precision of $0.84$ 
for a recall of $0.42$. In any case, as \Cref{fig:thr2precision} presents, our 
selection technique is significantly better than random, indicating it can save 
developer resources.

Among the selected bugs, we assess how effective the test rankings of \name are 
over a random baseline. The random approach randomly ranks the syntactic 
clusters (groups of syntactically equivalent FIB tests) of the generated tests.
We run the random baseline 100 times and average the results. 

Table~\ref{tab:ranking_performance} presents the ranking evaluation results. On 
the Defects4J benchmark, the ranking technique of \name improves upon the 
random baseline across all of the $acc@n$ metrics, presenting 30, 14, and 7 more BRTs than the random baseline on $n=1$, $3$, and $5$ respectively.
Regarding $acc@1$, the first column shows 
that 43\% of the top ranked tests produced by \name successfully reproduce the 
original bug report on the first try. When $n$ increases to 5, BRTs can be 
found in 57\% of the selected bugs, or 80\% of all 
reproduced bugs. The conservative threshold choice here,  
emphasizes recall over precision. However, if the threshold is raised, 
the maximum precision can rise to 0.8 (for $Thr=10$, $n=5$).

The $wef@n_{agg}$ values are additionally reported by both summing and 
averaging the $wef@n$ of all (350) selected bugs. The summed $wef@n$ value 
indicates the total number of non-BRTs that would be manually examined within 
the top $n$ ranked tests. Smaller $wef@n$ values indicate that a technique
delivers more bug reproducing tests. Overall, the ranking of 
\name saves up to 14.5\% of wasted effort when compared to the random baseline, 
even after bugs are selected. Based on 
these results, we conclude that \name can reduce wasted inspection effort and 
thus be useful to assist developers.

\begin{table}[t]
    \caption{Ranking Performance Comparison between \name and Random Baseline\label{tab:ranking_performance}}
    \scalebox{0.74}{
    \setlength\tabcolsep{4.2pt}
    \begin{tabular}{@{}l|ll|ll|ll|ll@{}}
    \toprule
    & \multicolumn{4}{c|}{Defects4J}   & \multicolumn{4}{c}{GHRB}                                   \\ \midrule
    & \multicolumn{2}{c|}{$acc@n$ ($precision$)}     & \multicolumn{2}{c|}{$wef@n_{agg}$}     & \multicolumn{2}{c|}{$acc@n$ ($precision$)}     & \multicolumn{2}{c}{$wef@n_{agg}$}     \\ \midrule
    $n$ & \name & random & \name & random & \name & random & \name & random \\ \midrule
    $1$   & \textbf{149} (\textbf{0.43})   & 116 (0.33)  & \textbf{201} (\textbf{0.57})   & 234 (0.67) &  \textbf{6} (\textbf{0.29})  &  4.8 (0.23)   &  \textbf{15} (\textbf{0.71}) &  16.2 (0.77)   \\
    $3$   & \textbf{184} (\textbf{0.53})   & 172 (0.49)  & \textbf{539} (\textbf{1.54})   & 599 (1.71) &  \textbf{7} (\textbf{0.33})  &  6.6 (0.31)   &  \textbf{42} (\textbf{2.0})  &  44.6 (2.12)   \\
    $5$  & \textbf{199} (\textbf{0.57})    & 192 (0.55)  & \textbf{797} (\textbf{2.28})   & 874 (2.5) &   \textbf{8} (\textbf{0.38})  &  7.3 (0.35)  &   \textbf{60} (\textbf{2.86}) &  64.3 (3.06)    \\ \bottomrule
    \end{tabular}}
\end{table}

\begin{tcolorbox}[boxrule=0pt,frame hidden,sharp corners,enhanced,borderline north={1pt}{0pt}{black},borderline south={1pt}{0pt}{black},boxsep=2pt,left=2pt,right=2pt,top=2.5pt,bottom=2pt]
    \textbf{Answer to RQ2-3:} \name can reduce both
    the number of bugs and tests that must be inspected: 33\% of the bugs are safely discarded while preserving 87\% of the successful bug reproduction. Among selected bug sets, 80\% of all bug reproductions can be found within 5 inspections. 
\end{tcolorbox}  
\subsection{RQ3. How well would \name work in practice?}

\begin{table}[ht]
    \centering
    \caption{Bug Reproduction in GHRB:
    x/y means x reproduced out of y bugs\label{tab:dhr_performance}}
    \scalebox{0.94}{
    \begin{tabular}{lr|lr|lr}
    \toprule
    Project & rep/total & Project & rep/total & Project & rep/total\\\midrule
    AssertJ & 3/5 & Jackson & 0/2 & Gson & 4/7 \\
    checkstyle & 0/13 & Jsoup & 2/2 & sslcontext & 1/2 \\
    \bottomrule
    \end{tabular}
    }
\end{table} 

\subsubsection{RQ3-1} We explore the performance
of \name when operating on the GHRB dataset of recent bug reports. 
We find that of the 31 bug reports we study, \name can automatically 
generate bug reproducing tests for 10 bugs based on 50 trials, for a 
success rate of \textbf{32.2\%}. This success rate is similar to the results
from Defects4J presented in RQ1-1, suggesting \name generalizes to new
bug reports. A breakdown of results by project is provided in 
\Cref{tab:dhr_performance}. Bugs are successfully reproduced in AssertJ, Jsoup, 
Gson, and sslcontext, while they were not reproduced in the other two. We could 
not reproduce bugs from the Checkstyle project, despite it having a large 
number of bugs; upon inspection, we find that this is because the project's 
tests rely heavily on external files, which \name has no access to, as shown in 
\Cref{sec:rq33_example}. \name also does not generate BRTs for the Jackson project,
but the small number of bugs in the Jackson
project make it difficult to draw conclusions from it.

\begin{tcolorbox}[boxrule=0pt,frame hidden,sharp corners,enhanced,borderline north={1pt}{0pt}{black},borderline south={1pt}{0pt}{black},boxsep=2pt,left=2pt,right=2pt,top=2.5pt,bottom=2pt]
    \textbf{Answer to RQ3-1:} \name is capable of generating bug reproducing
    tests even for recent data, suggesting it is not simply remembering 
    what it trained with. % without possibility of training data contamination.
\end{tcolorbox}  

\subsubsection{RQ3-2} \name uses several predictive factors correlated with successful bug reproduction for selecting bugs and ranking tests. In this research question, we check whether the identified patterns based on the Defects4J dataset continue to hold in the recent GHRB dataset. 

\begin{figure}[ht]
    \centering
    \begin{subfigure}{0.24\textwidth}
        \centering
        \includegraphics[width=\linewidth]{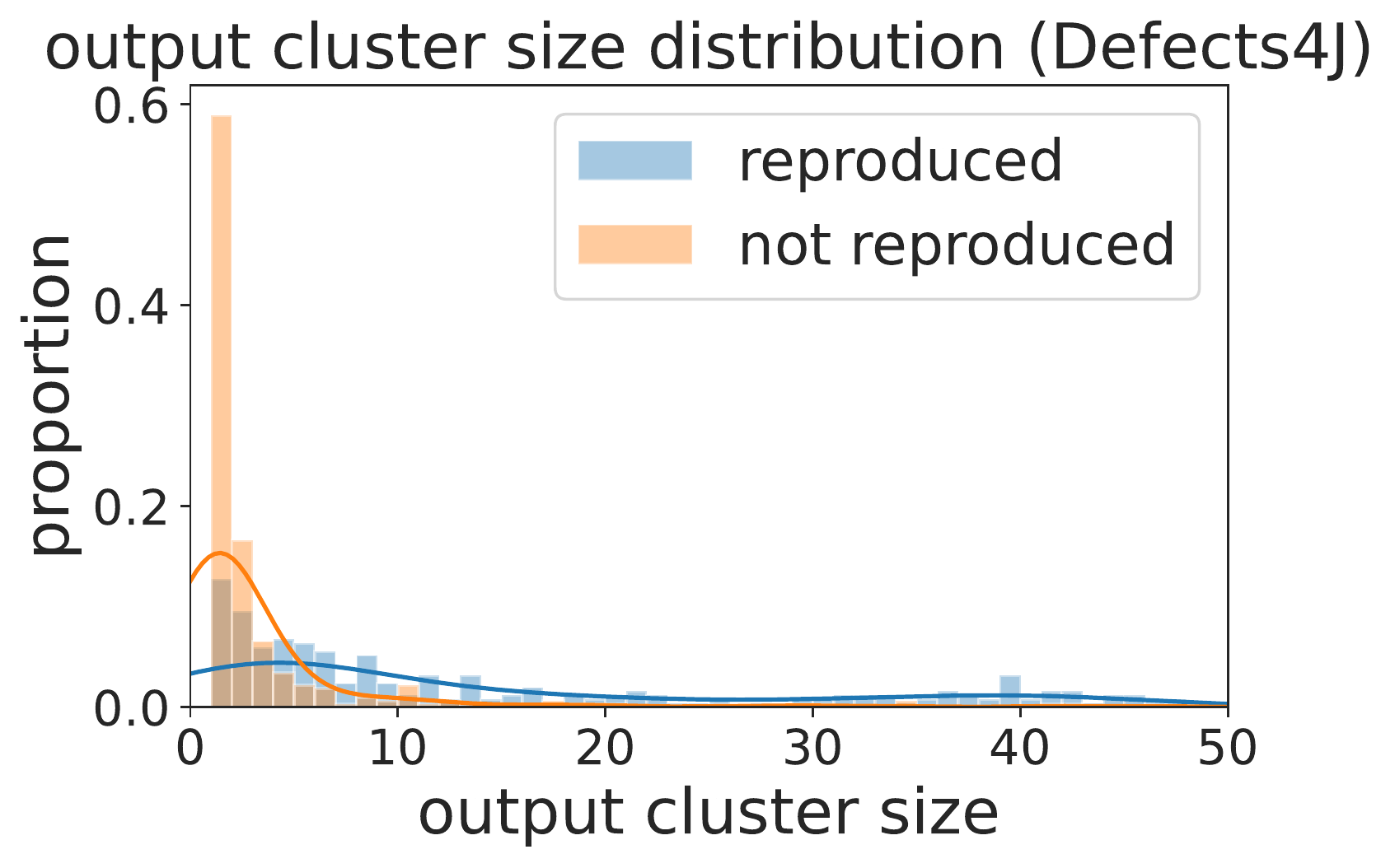}
    \end{subfigure}%
    \begin{subfigure}{0.24\textwidth}
        \centering
        \includegraphics[width=\linewidth]{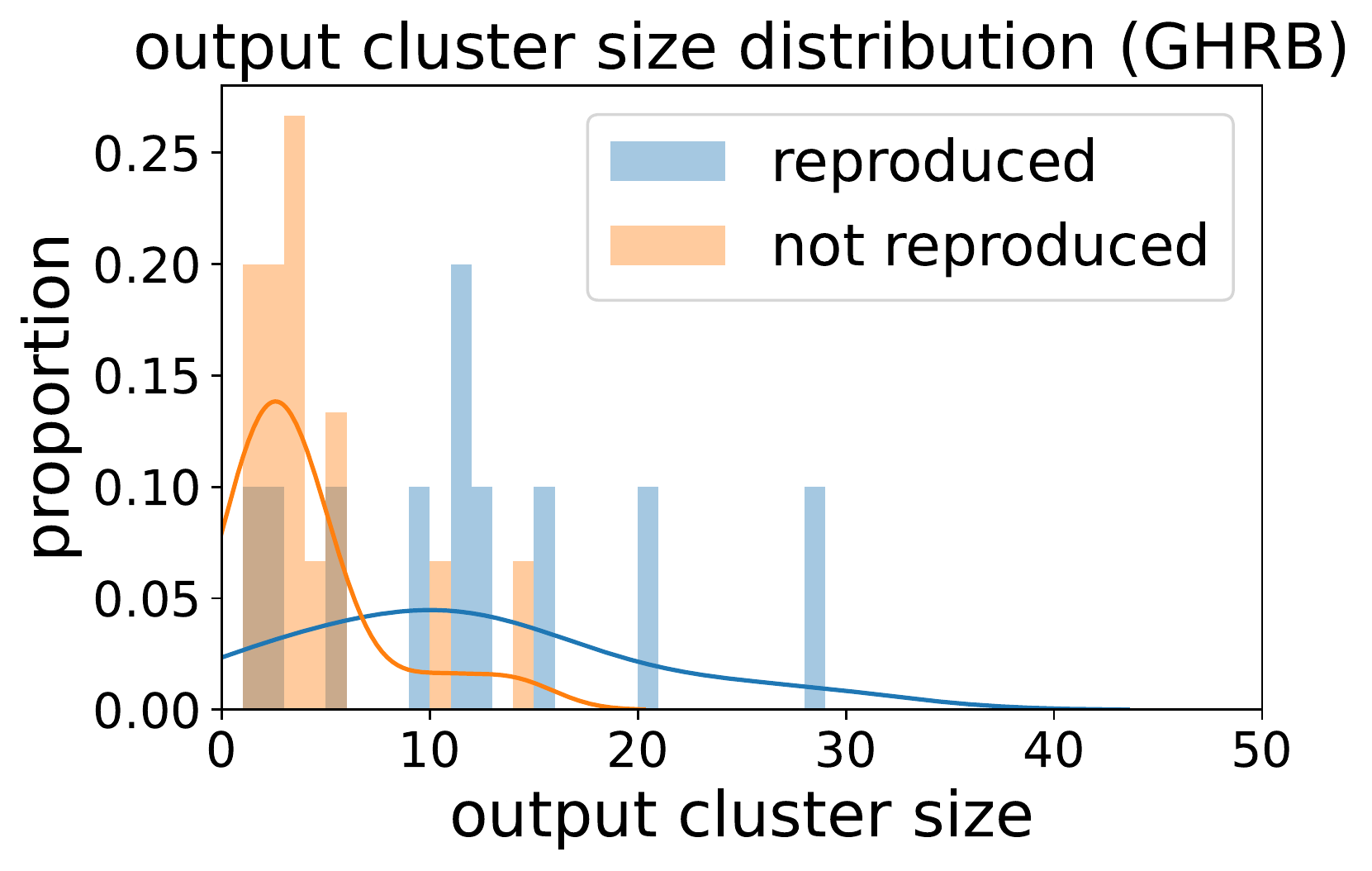}
    \end{subfigure}
    \caption{Distribution of the $max\_output\_clus\_size$ values for reproduced and not-reproduced bugs}
    \label{fig:output_cluster_size_pattern}
\end{figure}

Recall that we use the maximum output cluster size as a measure of agreement among the FIBs, and thus as a selection 
criterion to identify whether a bug has been reproduced. 
To observe whether the criterion is a reliable indicator to 
predict the success of bug reproduction, we observe the trend of $max\_output\_clus\_size$ between the 
two datasets, with and without BRTs. In Figure~\ref{fig:output_cluster_size_pattern}, we see that 
the bugs with no BRT typically have small $max\_output\_clus\_size$, mostly under ten; this pattern is consistent in both datasets.

The ranking results of GHRB are also presented in Table~\ref{tab:ranking_performance}. They are consistent to the results from Defects4J, indicating the features used for our ranking strategy continue to be good indicators of successful bug reproduction.

\begin{tcolorbox}[boxrule=0pt,frame hidden,sharp corners,enhanced,borderline north={1pt}{0pt}{black},borderline south={1pt}{0pt}{black},boxsep=2pt,left=2pt,right=2pt,top=2.5pt,bottom=2pt]
    \textbf{Answer to RQ3-2:} We find that the factors used for the ranking and selection of \name consistently predict bug reproduction in real-world data.
\end{tcolorbox}  

\subsubsection{RQ3-3} 
\label{sec:rq33_example}
We present case studies of attempts by \name to 
reproduce bugs that either succeeded or failed.

\begin{table}[h!]
    \caption{\label{tab:assertj-success} Bug Report Successfully Reproduced: URLs are omitted for brevity (AssertJ-Core Issue \#2666)}
    \begin{tabular}{@{}lp{0.4\textwidth}@{}}
    \toprule
    \textbf{Title}     & \textbf{assertContainsIgnoringCase fails to compare i and I in tr\_TR locale} \\ \midrule
    \multicolumn{2}{l}{\begin{tabular}[c]{@{}l@{}} See org.assertj.core.internal.Strings\#assertContainsIgnoringCase \\

        [url] \\

        I would suggest adding [url] verification to just ban \\
        toLowerCase(), toUpperCase() and other unsafe methods: \#2664\end{tabular}} \\ \bottomrule
    \end{tabular}
\end{table}

\begin{lstlisting}[basicstyle=\footnotesize\ttfamily,
    float=t,
    columns=flexible,
    breaklines=true,
    language=java,
    caption={Generated FIB test for AssertJ-Core-2666.},
    label={lst:assertj-succ-test},]
public void testIssue952() {
  Locale locale = new Locale("tr", "TR");
  Locale.setDefault(locale);
  assertThat("I").as("Checking in tr_TR locale").containsIgnoringCase("i");
}
\end{lstlisting}

We first present a successfully reproduced bug report, issue \#2685 for 
the AssertJ-Core project, in \Cref{tab:assertj-success}. The bug is that, in \code{tr\_TR} 
locale, \code{assertContainsIgnoringCase} method is mishandling the letter 
\code{I}. For this bug report, \name generates the bug reproducing test in 
Listing~\ref{lst:assertj-succ-test}. Despite the bug report not containing any 
executable code, \name successfully generated a bug reproducing test that 
compares \code{I} and \code{i}, using the \code{containsIgnoringCase} method 
referenced within the report. A BRT for this bug is ranked at the second place, 
suggesting a developer would have been quickly provided with a reproducing test.

\begin{table}[h!]
    \caption{\label{tab:checkstyle-failure} Bug Report Reproduction Failure: Lightly edited for clarity (Checkstyle Issue \#11365)}
    \begin{tabular}{@{}lp{0.4\textwidth}@{}}
    \toprule
    \textbf{Title}     & \textbf{FinalClassCheck: False positive with anonymous classes} \\ \midrule
    \multicolumn{2}{l}{\begin{tabular}[c]{@{}l@{}} ... I have executed the cli and showed it below, as cli describes the \\
        problem better than 1,000 words\\
        $\rightarrow$\code{src cat Test.java} \\
        \code{[...]} \\
        \code{public class Test \{} \\
        \hspace*{1em}\code{class a \{ // expected no violation} \\
        \hspace*{2em}\code{private a() \{\} \} \}} \\
        \code{[...]} \\
        $\rightarrow$\code{java [...] -c config.xml Test.java} \\
        \code{Starting audit...} \\
        \code{[ERROR] Test.java:3:5: Class a should be declared as final.} \\
        % \code{Class Fan-Out Complexity is 3 (max allowed is 1).}\\
        \end{tabular}} \\ \bottomrule
    \end{tabular}
\end{table}

\begin{lstlisting}[basicstyle=\footnotesize\ttfamily,
    columns=flexible,
    breaklines=true,
    language=java,
    caption={Generated FIB test for Checkstyle-11365.},
    label={lst:checkstyle-fail-test},]
public void testFinalClass() throws Exception {
    final DefaultConfiguration checkConfig =
            createModuleConfig(FinalClassCheck.class);
    final String[] expected = CommonUtil.EMPTY_STRING_ARRAY;
    verify(checkConfig, getPath("InputFinalClassAnonymousClass.java"), expected);
}
\end{lstlisting}

We now turn to a bug report that is not successfully reproduced. 
\Cref{tab:checkstyle-failure} contains issue \#11365 in the Checkstyle project. 
The bug is that CheckStyle incorrectly decides that a class should be declared 
\code{final}, and mistakenly raises an error. A FIB test generated by \name is presented in 
Listing~\ref{lst:checkstyle-fail-test}, which fails as the Java file it 
references in Line 5 is nonexistent. This highlights a weakness of \name, 
i.e., its inability to create working environments outside of source code for the generated tests. 
However, if we put the content of \texttt{Test.java} from the report into the 
referenced file, the test successfully reproduces the bug, indicating that the 
test itself is functional, and that even when a test is initially
incorrect, it may reduce the amount of developer effort that goes into writing 
reproducing tests.

\section{Discussion}

\subsection{Manual Analysis of \name Failures}

Despite successfully reproducing 33.5\% of the Defects4J bugs,
in many cases \name could not reproduce the bugs from the bug reports.
To investigate which factors may have caused \name to struggle, we
manually analyzed 40 bug reports and corresponding \name outputs. The
most common cause of failure, happening in 13 cases, was \emph{a need of helper definitions}:
while the developer-written tests made use of custom testing helper
functions which at times spanned hundreds of lines, 
\name-generated tests were generally agnostic to such functions,
and as a result could not adequately use them. This points to a need to
incorporate project-specific information for language models to further
improve performance. Other failure reasons included low report quality in 11 cases
(i.e., a human would have difficulty reproducing the issue as well),
the LLM misidentifying the expected behavior in 8 cases, dependency on
external resources in 6 cases (as was the case in Listing~\ref{lst:checkstyle-fail-test}),
and finally insufficient LLM synthesis length in 3 cases. This taxonomy of
failures suggests future directions to improve \name, which we hope to explore.

\subsection{Code Overlap with Bug Report}
As Just et al. point out~\cite{just2018comparing}, bug reports can already contain partially or fully executable test code, but developers rarely adopt the provided tests as is.
To investigate whether \name relies on efficient extraction of report content or effective
synthesis of test code,
we analyzed the 750 bug reports from Defects4J used in our experiment. We find that 19.3\% of them 
had full code snippets (i.e., code parsable to a class or method), while 39.2\% had partial code 
snippets (i.e., not a complete class or method but in the form of source code statements or 
expressions); finally 41.5\% did not contain code snippets inside. 
Considering only the 251 bug reports that 
\name successfully reproduced, the portion of containing the
full snippets got slightly higher (25.1\%), whereas the portion of bug reports with partial
snippets was 37.9\%, and 37.1\% did not have code snippets. When \name generated tests from
bug reports containing any code snippets, we find that on average 81\% of the tokens in the body of
the \name-generated test methods overlapped with the tokens in the code snippets.

We note that using full code snippets provided in reports does not always reproduce the bug 
successfully; recall that the Copy\&Paste baseline in \Cref{fig:baseline-venn} succeeded only on 
36 bugs. Although whether a bug report contains full code snippets or not may affect the success 
or failure of \name, \name generated correct bug reproducing tests even from incomplete code, or 
without any code snippets. Thus, we argue that \name is capable of both extracting relevant code 
elements in bug report and synthesizing code aligned with given a natural language description.

\section{Threats to Validity}
\label{sec:threats}

\noindent\textbf{Internal Validity} concerns whether our experiments
demonstrate causality. In our case, two sources 
of randomness threat internal validity: the flakiness of tests and the
randomness of LLM querying. While we do observe a small number
of flaky tests generated, the number of them is significantly smaller (<2\%) than
the overall number of tests generated, and as such we do not believe their existence
significantly affects our conclusions. Meanwhile, we engage with the randomness of the
LLM, performing an analysis in RQ2-1.

\noindent\textbf{External Validity} concerns whether the results presented would
generalize. In this case, it is difficult to tell whether the results we 
presented here would generalize to other Java projects, or projects in other 
languages. 
While the uniqueness of our prompts and our use of GHRB cases provide 
some evidence that \name is not simply 
relying on the memorization of the underlying LLM, it is true that \name 
benefits from the fact that the underlying LLM, Codex, has likely seen the 
studied Defects4J projects during training. However, our aim is \emph{not} to 
assess whether a specific instance of Codex has general intelligence about 
testing: our aim is to investigate the extent to which LLM architectures 
augmented with post-processing steps can be applied to the task of bug 
reproduction. For \name to be used for an arbitrary project with a similar 
level of efficacy as in our study, we expect the LLM of \name to have 
seen projects in a similar domain, or the target project itself. 
This can be achieved with fine-tuning the LLMs, as 
studied in other domains~\cite{Tinn2021xy,Wang2021qv} (note that Codex is GPT-3 
fine tuned with source code data). As a due diligence, we checked 
how many tests generated from the Defects4J benchmark verbatim matched 
developer-committed bug reproducing tests. There were only such three cases, 
and all had the test code written verbatim in the report as well, suggesting it 
is likely they got verbatim answers from the report rather than from 
memorization. 
We also report a few general 
conditions for which \name does not perform well: it does not generalize to 
tests that rely on external files or testing infrastructure whose syntactic 
structure is significantly different from the typical JUnit tests (such as the 
Closure project in Defects4J).

\section{Related Work}
\label{sec:relwork}

\subsection{Test Generation}
Automated test 
generation has been explored since almost 50 years ago~\cite{Miller:1976ht}. 
The advent of the object-oriented programming paradigm caused a shift in test 
input generation techniques, moving from primitive value exploration to 
deriving method sequences to maximize coverage~\cite{Pacheco:2007oq,
Fraser:2013vn}. However, a critical issue with these techniques is the oracle 
problem~\cite{Barr:2015qd}: as it is difficult to determine what the correct 
behavior for a test should be, automated test generation techniques 
either rely on implicit oracles~\cite{Pacheco:2007oq}, or accept the 
current behavior as correct - effectively generating regression 
tests~\cite{Fraser:2013vn,Tufano2020ji}. Swami~\cite{Motwani2019Swami} generates
edge-case tests by analyzing structured specifications using rule-based heuristics; 
its ``rigid''ness causes it to fail when the structure deviates from
its assumptions, whereas \name makes no assumptions on bug report structure.

Similar to our work, some techniques focus on reproducing 
problems reported by users: a commonly used \emph{implicit} oracle is that the 
program should not crash~\cite{Barr:2015qd}. Most of existing work~\cite{Nayrolles2015Jcharming, 
chen2014star, xuan2015crash, Soltani2018aa, Derakhshanfar2020wt} aim to reproduce crashes given a 
stack trace, which is assumed to be provided by a user. 
Yakusu~\cite{Fazzini2018aa} and ReCDroid~\cite{Zhao2019na}, on the other hand, require user reports written in specific formats 
to generate a crash-reproducing test for mobile applications. All the
crash-reproducing techniques differ significantly from our work as they rely on 
the crash-based implicit oracle, and make extensive use of of SUT code (i.e., they are white-box techniques). BEE~\cite{Song2020Bee} automatically parses bug 
reports to classify sentences that describe observed or expected behavior but 
stops short of actually generating tests.
To the best of our knowledge, we are the first to propose a technique to 
reproduce general bug reports in Java projects.

\subsection{Code Synthesis}
Code synthesis also has a long history of research. Traditionally, code 
synthesis has been approached via SMT solvers in the context of Syntax-Guided 
Synthesis (SyGuS)~\cite{Alur2015SyGuS}. As machine learning techniques 
improved, they showed good performance on the code synthesis task; 
Codex demonstrated that an LLM could solve programming tasks based on natural 
language descriptions~\cite{Chen2021ec}. Following Codex, some found that 
synthesizing tests along with code was useful: AlphaCode used automatically 
generated tests to boost their code synthesis performance~\cite{Li2022kp}, 
while CodeT jointly generated tests and code from a natural language 
description~\cite{Chen2022iw}. The focus of these techniques is not on test 
generation; on the other hand, \name processes LLM output to maximize the 
probability of execution, and focuses on selecting/ranking tests to reduce the 
developer's cognitive load.

\section{Conclusion}
\label{sec:conclusion}

In this paper, we first establish that the report-to-test problem is
important, by inspecting relevant literature and performing an analysis on 300 
open source repositories. To solve this problem, we introduce \name, a 
technique that uses a pretrained LLM to analyze bug reports, generate
prospective tests, and finally rank and suggest the generated solutions based 
on a number of simple statistics. Upon extensive analysis, we find that \name 
is capable of reproducing a significant number of bugs in the Defects4J 
benchmark, and perform detailed analyses about the requirements of 
using the technique. We further experiment with a real-world report-to-bug
dataset that we have collected: we find that \name shows similar performance on 
this dataset when compared to the Defects4J benchmark, demonstrating its versatility. 
In both datasets, \name successfully identifies
when the bug is reproduced by which test, showing
that \name can minimize developer effort as well.
We hope to expand upon these results and explore the synergy with
existing test-generation techniques to further help practitioners.

\section*{Acknowledgment}

This work was supported by the National Research Foundation of Korea (NRF) Grant
(NRF-2020R1A2C1013629), 
Engineering Research Center Program through the National Research Foundation of Korea (NRF) funded by the
Korean Government MSIT (NRF-2018R1A5A1059921), and the
Institute for Information \& Communications Technology 
Promotion grant funded by the Korean government MSIT (No.2022-0-00995).

\bibliographystyle{IEEEtran}
\bibliography{extracted}
\end{document}